\documentclass[preprint,12pt]{elsarticle}
\biboptions{sort&compress,longnamesfirst,numbers}
\usepackage[english]{babel}
\usepackage{geometry}
\usepackage{lineno}
\usepackage{multirow}
\usepackage{graphicx}
\usepackage{tabularx}
\usepackage{mathtools}
\usepackage{threeparttable}
 \newcolumntype{L}{>{\raggedright\arraybackslash}X}
\usepackage{booktabs}
\usepackage{soul}
\usepackage[export]{adjustbox}
\usepackage{subfig}
\usepackage{amsmath}
\usepackage{amssymb}
\usepackage{multirow}
\usepackage[titletoc]{appendix}

\usepackage{colortbl}
\usepackage{url}

\usepackage{breakurl}
\usepackage[breaklinks]{hyperref}
\definecolor{kugray5}{RGB}{224,224,224}
\makeatletter 
\renewcommand\@biblabel[1]{#1.} 
\makeatother
\usepackage{floatrow}
\usepackage[justification=centering]{caption}

\graphicspath{{Figures/}}

\addto\captionsenglish{}
\usepackage[markup=underlined]{changes}

\definechangesauthor[color=red]{DP}
\definechangesauthor[color=blue]{EB}
\definechangesauthor[color=blue]{MP}
\definechangesauthor[color=orange]{APB}
\usepackage{todonotes}


\makeatletter
\def\ps@pprintTitle{%
 \let\@oddhead\@empty
 \let\@evenhead\@empty
 \def\@oddfoot{}%
 \let\@evenfoot\@oddfoot}
\makeatother
\begin{document}

\begin{frontmatter}
\title{Assessing strategies to manage distributed photovoltaics in Swiss low-voltage networks: An analysis of curtailment, export tariffs, and resource sharing}
\author{Gerard Marias Gonzalez}
\author{Nicolas Wyrsch}
\author{Christophe Ballif}
\author{Alejandro Pena-Bello\corref{cor1}}
\cortext[cor1]{Corresponding author. E-mail: alejandro.penabello@epfl.ch}

\address{Photovoltaics and thin film electronics laboratory (PV-LAB), École Polytechnique Fédérale de Lausanne (EPFL), Institute of Electrical and Microengineering (IEM), Neuchâtel, Switzerland}

\begin{abstract}
The integration of photovoltaic systems poses several challenges for the distribution grid, mainly due to the infrastructure not being designed to handle the upstream flow and being dimensioned for consumption only, potentially leading to reliability and stability issues. This study investigates the use of capacity-based tariffs, export tariffs, and curtailment policies to reduce negative grid impacts without hampering PV deployment. We analyze the effect of such export tariffs on three typical Swiss low-voltage networks (rural, semi-urban, and urban), using power flow analysis to evaluate the power exchanges at the transformer station, as well as line overloading and voltage violations. Finally, a simple case of mutualization of resources is analyzed to assess its potential contribution to relieving network constraints and the economic costs of managing LV networks. We found that the tariff with capacity-based components on the export (CT export daily) severely penalizes PV penetration. This applies to other tariffs as well (e.g. IRR monthly, Curtailment 30, and DT variable) but to a lesser extent. However, the inclusion of curtailment at 50\% and 70\%, as well as mixed tariffs with capacity-based components at import and curtailment, allow for a high degree of PV installations in the three zones studied and help to mitigate the impact of PV on the distributed network.

\end{abstract}
\end{frontmatter}
\renewcommand{\thefootnote}{\alph{footnote}}

\begin{table}[h]
\centering
\scalebox{0.8}{
\begin{tabular}{llll}
PV & Photovoltaic & FiT & Feed-in Tariff\\
SC & Self-consumption & SS & Self-sufficiency\\
FT & Flat tariff & DT & Double tariff\\
CT & Capacity-based tariff & LV & Low-voltage\\
TOU & Time of use & MV & Medium-voltage\\
EV & Electric vehicle & HP & Heat pump\\
CAPEX & Capital expenditure & OPEX & Operational expenditure \\
TSO & Transmission system operator & DSO & Distribution system operator \\
SIA & Swiss Society of Engineers and Architects & IQR & Interquartile range\\
RegBL & Registre fédéral des bâtiments et des logements  & MILP & Mixed Integer Linear Programming\\
IRENA & International Renewable Energy Agency\\

\end{tabular}}
\caption*{List of abbreviations.}
\end{table}

\section{Introduction}\label{introduction}
The Swiss energy transition foresees 35 TWh of PV production by 2050 \cite{Mantelerlass_2023}, that is, a minimum of 30 GW of PV capacity is anticipated to be installed across the country. Rooftops are expected to host most of the PV installations, as the availability of land for solar farms is limited due to various factors such as land scarcity, citizen opposition, and/or legal requirements \cite{walch2022spatio}. However, the integration of distributed photovoltaics (PV) poses several challenges for the distribution grid, including over-voltage, line, and transformer overloading, and reverse power flow. These problems arise mainly because the distribution grid infrastructure was not designed to handle the upstream flow and was dimensioned for consumption only. As a result, the massive integration of PV systems may cause reliability and stability issues to the grid, affecting the adequate maintenance of the infrastructure and leading to inefficiencies not previously seen in the electricity industry.

Major issues on distribution networks due to high PV penetration arise mainly in rural grids and involve voltage violations and reverse power flow \cite{allard2020considering,pena2023balancing}. Effective management of these challenges is critical to ensure the reliability and stability of the power grid. Moreover, alternative tariffs were found to be able to drive the adoption of storage and recover the grid costs without causing relevant economic differences for the customers in Low-Voltage (LV) networks\cite{pena2023balancing}. However, they only marginally mitigated the impact of PV on the network since the main problems for the distribution grid, in general, stem from PV injection. In this sense, here we propose to explore solutions involving the management of PV export. This study aims to analyze PV curtailment, as well as its combination with capacity-based tariffs for import, capacity-based tariffs for export, and a diversity of export prices, including a block rate tariff, to mitigate the impact on the network during the crucial hours of the year and, in this way, help to relieve transformer overloading, without hampering the economic attractiveness of potential rooftop PV installations to reach the national PV targets. Furthermore, with the increasing interest in peer-to-peer trading and its potential contribution to relieving network constraints and the environmental and economic costs of managing the LV networks, we aim to analyze a single case of mutualization of resources.

The different scenarios are tested in three typical Swiss LV networks (a rural network, a semi-urban network, and an urban network), using power flow analysis to evaluate the power exchanges at the transformer station, as well as line overloading, and voltage violations. We look at the year 2025, using the expected cost of the PV and battery systems for such year \cite{irena_esr2017}, that are assumed to be installed at once for all buildings, we use real data from Swiss households based on one full year of demand data with a 15-minute resolution collected back in 2015 from a different case study \cite{Flexi1_2015, Flexi2_2017}, that is then adjusted to the transformer present-day load at the transformer level \cite{holweger2021privacy} (see Materials and methods).

We analyze the impact on the LV networks taking into account the PV hosting capacity, PV curtailment, amount of storage installed, load duration curves at the transformers, permissible overload for varying periods, line loading, and voltage deviations.

\section{Materials and methods}
The materials and methods applied in this study are consistent with those described in the previous publication \cite{pena2023balancing}, with the exception of the modifications outlined in Sections \ref{tar} and \ref{lab:mutua}.
\subsection{Input data}
\subsubsection{Demand and PV data}
Since demand data measurements are unavailable for the selected LV networks, an estimate of building demand is required. To achieve this, data is gathered from three sources to estimate load curves for each building in each zone. Firstly, information from the \textit{Registre fédéral des bâtiments et des logements} (RegBL) is used, including building date of construction or renovation, building or local categories, building surface, dwellings surface, and the number of floors. Secondly, data from the Swiss Society of Engineers and Architects (SIA) norms is utilized, which provides electrical consumption per building type and age. Thirdly, smart-meter measurements from the Swiss French-speaking area collected from the Flexi projects, which have a 15-minute resolution for different building categories, including apartments, houses, and non-residential buildings \cite{holweger2021privacy, Flexi1_2015, Flexi2_2017}. These measurements are used as a proxy variable. Finally, the real load curve at the transformer station, provided by Romande Energie, is utilized to match the sum of individual building loads at the aggregated level.

Moreover, data on roof surface area, azimuth, and tilt for each building in the selected LV network is extracted from the RegBL. To model PV generation, monitored outdoor temperature and horizontal solar irradiance are obtained from the MeteoSwiss weather station located in Pully, Vaud. Data from the Sandia PV module database for the SunPower SPR-315E-WHT PV modules, which are monocrystalline silicon with an efficiency of 19.3\% and a module area of 1.6310 $m^{2}$, are combined with the monitored environmental variables to calculate the current and voltage performance per module. The modules are then aggregated depending on the roof orientation.

Finally, PV and battery price levels for the reference year 2025 are determined from data in the IRENA report \cite{irena_esr2017}, calibrated to Swiss price levels in a market study  \cite{suisseenergie2021} using the empirical model proposed by Bloch et al.  \cite{bloch2019impact}. The resulting PV and battery cost function is separated into a fixed and a linear component. The fixed component includes administrative procedures and fixed installation costs, while the linear component is proportional to the installed capacity. These linear cost functions are annualized and combined with the PV yearly maintenance cost and battery lifetime in Eq. \ref{eq:obj}. Table \ref{tab:param} shows the main techno-economic parameters of this study.

\begin{table}[]
\begin{tabular}{lll}
\toprule
Parameter              & Value  &    Reference  \\\midrule
System lifetime        & 25 years  &  \cite{irena_esr2017} \\
Discount rate          & 3\%        & Own assumption\\
PV fixed cost          & 10 049 \,€ &  \cite{irena_esr2017,suisseenergie2021}\\
PV specific costs      & 1.05 \,€/W &  \cite{irena_esr2017,suisseenergie2021}\\
Battery fixed cost     & 0 \,€      &  \cite{irena_esr2017,suisseenergie2021}\\
Battery specific costs & 229 \,€/kWh & \cite{irena_esr2017,suisseenergie2021}\\\bottomrule
\end{tabular}
\caption{Techno-economic parameters used in this study.}
\label{tab:param}

\end{table}

\subsubsection{Low-voltage network data}
\begin{figure}[]
    
\newcommand{\fhe}{4.2cm}
    \centering

    \subfloat[\label{fig:grids_urban} Urban LV network.]{\includegraphics[width=0.5\textwidth]{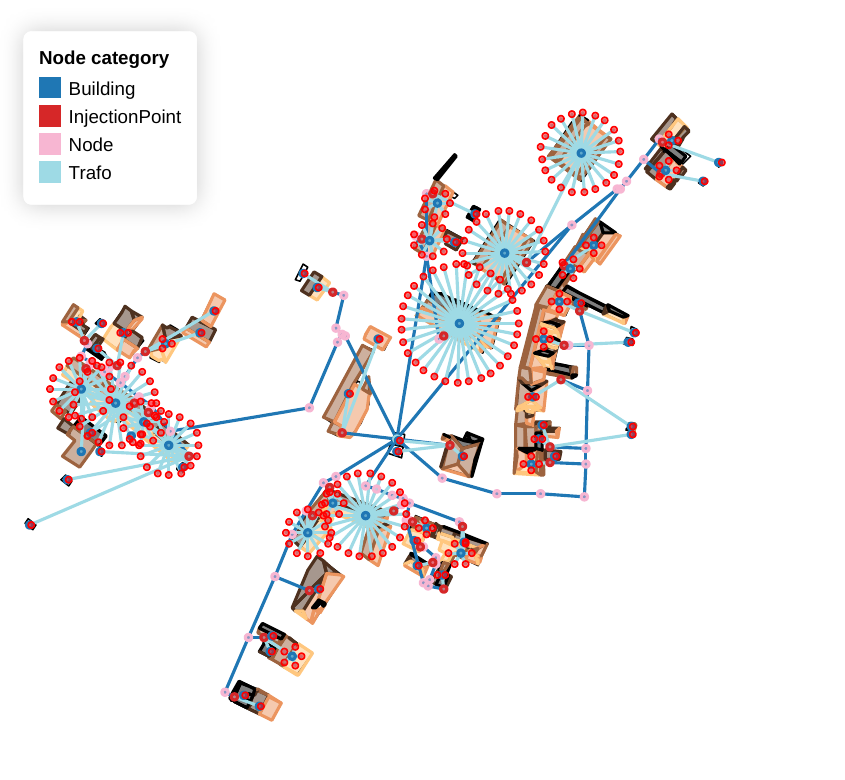}} 
    \subfloat[\label{fig:grids_semi-urban}Semi-urban LV network.]{ \includegraphics[width=0.5\textwidth]{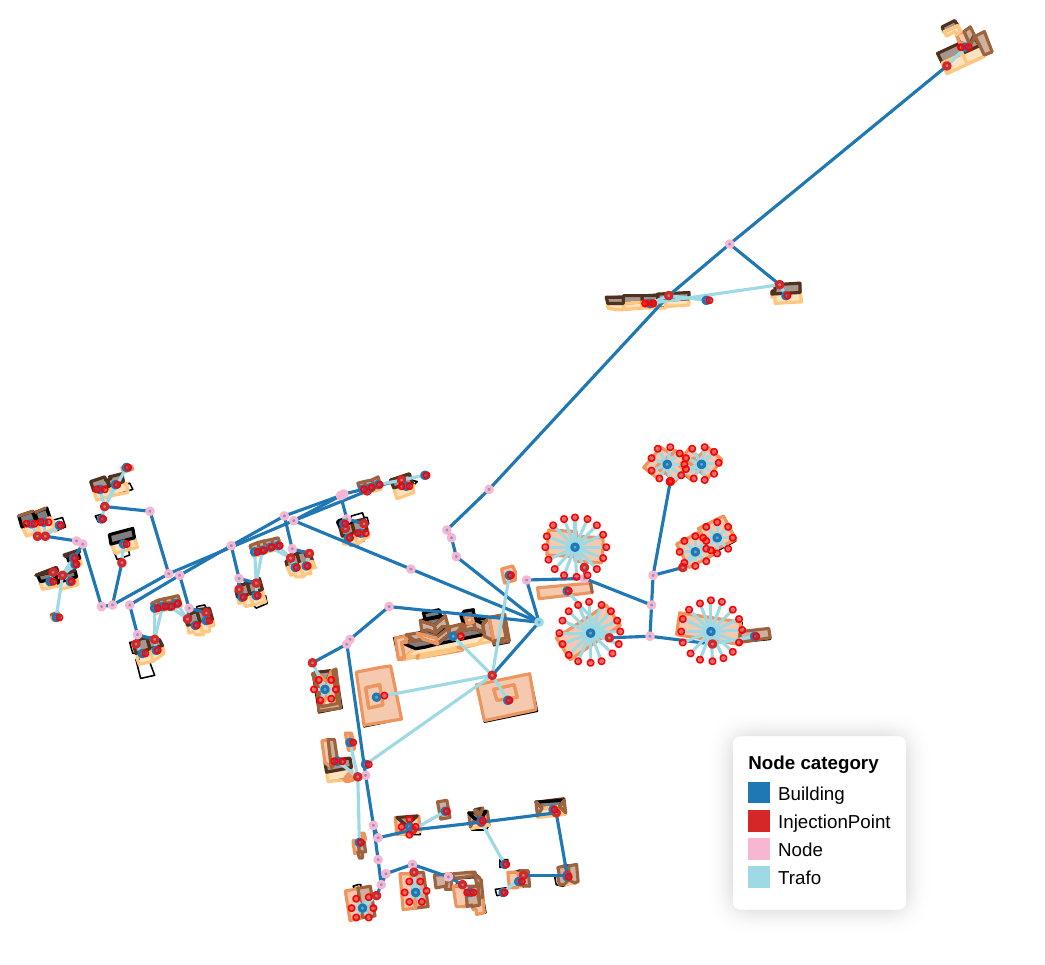}}\\
    
    \subfloat[\label{fig:grids_rural} Rural LV network.]{\includegraphics[width=0.5\textwidth]{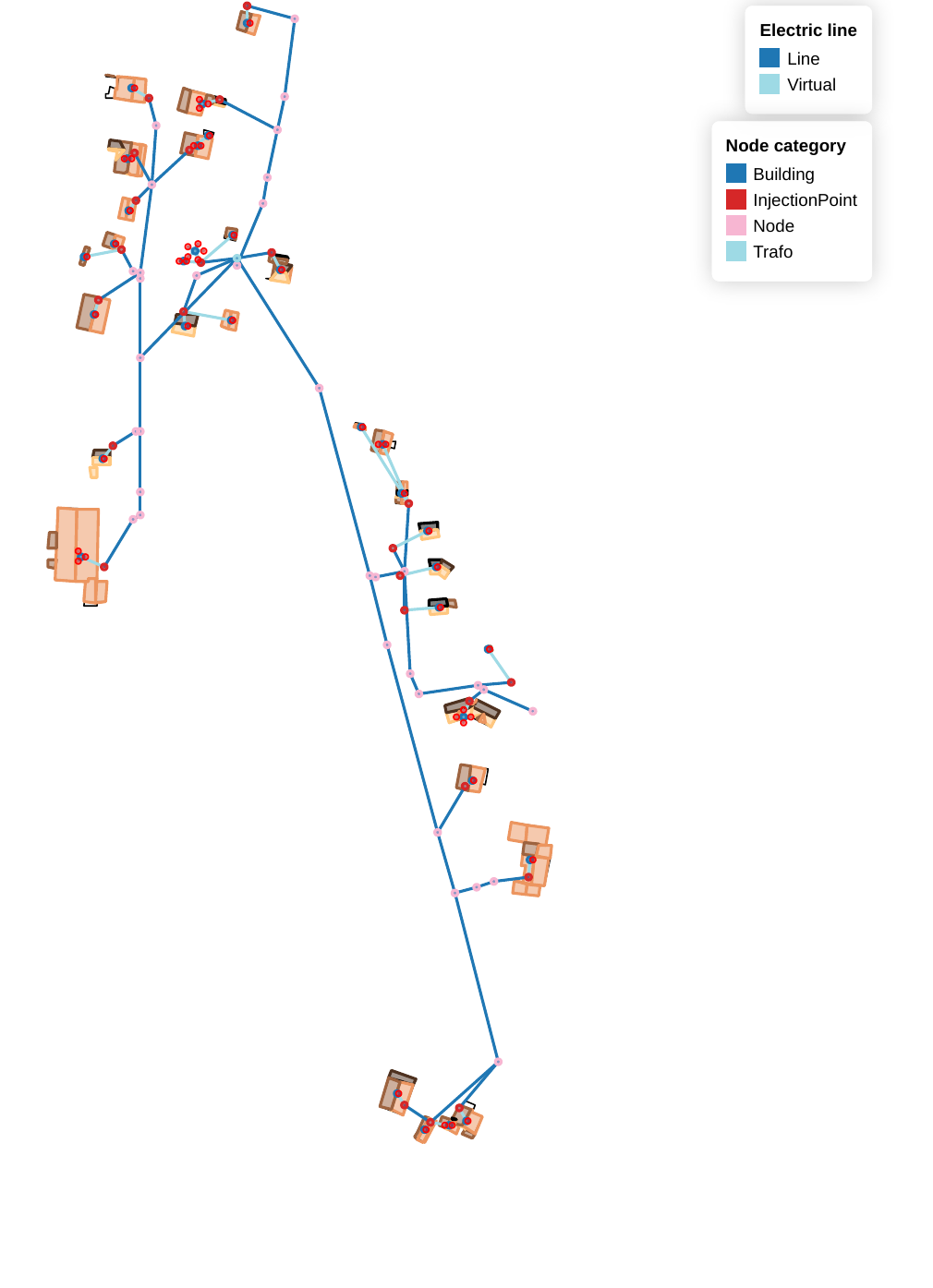}}
    \caption{Illustration of the LV networks used in this study.}
    \label{fig:grids}
\end{figure}

For each LV network, Romande Energie furnished details on the transformer, cables, and line characteristics, such as capacity, line length, and impedance. Additionally, the federal building identifier (EGID) of each customer linked to each injection point and the measured load charge of the transformer station, including correction for existing PV capacity, were also provided. The three LV networks were selected to represent rural, semi-urban, and urban LV networks, as displayed in Figure \ref{fig:grids}. The main features of each LV network are presented in Table \ref{tab:grids}.

\begin{table}[]
\scalebox{0.65}{
\begin{tabular}{llllllllll}
\toprule
LV network type                   & \multicolumn{1}{{m{2.cm}}}{Transformer capacity [kVA]} & \multicolumn{1}{m{2.5cm}}{Virtual Transformer Capacity [kVA]} & 
\multicolumn{1}{{m{2.cm}}}{Number of loads} & 
\multicolumn{1}{{m{2.3cm}}}{Number of buildings} & 
\multicolumn{1}{{m{2.cm}}}{Number of injection points} & 
\multicolumn{1}{{m{2.cm}}}{Max power [kW]} & 
\multicolumn{1}{{m{2.5cm}}}{PV correction max [kW]} & 
\multicolumn{1}{{m{2.5cm}}}{Total consumption [MWh]} \\\midrule
Urban                 & 2x 630                     & 1000                               & 287              & 65                   & 33                          & 419.8               &               -         & 1671.2               \\
Semiurban                  & 2x 400                     & 1000                               & 174              & 71                   & 37                          & 508.6              &           -             & 2065.6             \\
Rural & 1x 630                     & 630                                & 48               & 32                   & 24                          & 75.9                & 59.7                   & 204.3 \\\bottomrule
\end{tabular}}
\caption{Main characteristics of the low-voltage networks.}
\label{tab:grids}

\end{table}
\subsection{Tariffs}\label{tar}
Concerning the previous study \cite{pena2023balancing} where we analyzed only the import tariffs, here we focus on the export tariffs as well as on different amounts of curtailment and combinations of tariffs (e.g. volumetric export price, curtailment, and a combination of volumetric tariffs and capacity-based tariffs at the import).

This study evaluates seven different export price and tariff structures. All scenarios are compared against a reference import tariff, specifically the double tariff (reference DT) from Romande Energie (2022). The scenarios explore five different combinations for export pricing. First, purely volumetric export prices. Second, a combination of volumetric export prices and curtailment at various thresholds (30\%, 50\%, and 70\%). Third, a combination of volumetric export prices and a capacity-based tariff applied to injections (charged to the client) to mitigate PV feed-in peaks. Fourth, a combination of volumetric export prices, capacity-based tariffs at the import, and curtailment at different thresholds. Finally, an export price structured as a block rate tariff is also considered. Table \ref{tab:tariffs} presents the main components of each tariff, which are described below.

\begin{enumerate}
    \item Variable volumetric export price DT during the day (peak at 9.5 CHF/kWh, off-peak at 5 CHF/kWh): 9:00-13:59, from the 1st of April until August 31st (variable DT).
    \item An export price based on the monthly irradiance (IRR monthly), with a maximum of 14.05 CHF/kWh during the month with the lowest irradiance and a minimum of 3.6 CHF/kWh during the month with the highest irradiance.
    \item A flat export price of 9.5 CHF/kWh with a curtailment based on the installed capacity of the PV system at three different thresholds, namely, 30\%, 50\%, and 70\%. Across the curtailment scenarios presented (i.e. Curtailment 30\%, 50\%, and 70\%), the percentage given represents the highest amount of power that can be fed into the grid with the nominal capacity of the PV system. For instance, under a policy of Curtailment 30, the owner of a PV system of 10 kW, can only inject up to 3 kW into the grid.
    \item A combination of a flat export price of 9.5 CHF/kWh paid to the client and a capacity component \textbf{on the export} with a daily billing horizon (i.e. CT export daily  with the following prices:  [0.8116, 1.0108, 1.2400] CHF/$kW_{day-j}$) to be paid by the client according to the maximum power injected into the system.
    \item A combined instrument, including a flat export price of 9.5 CHF/kWh paid to the client and a capacity component with a daily billing horizon (i.e. CT daily, with the following prices:  [0.040, 0.070, 0.110] CHF/$kW_{day-j}$) to be paid by the client according to the maximum power drawn from the system and curtailment at 30\%, 50\%, and 70\% of the installed capacity of the PV system.
    \item A combined instrument, including a flat export price of 9.5 CHF/kWh paid to the client and a capacity component with a monthly billing horizon (i.e. CT monthly with a tariff of 0.7 CHF/$kW_{month-j}$) to be paid by the client according to the maximum power drawn from the system and curtailment at 30\%, 50\%, and 70\% of the installed capacity of the PV system.
    \item An export price based on a block rate with three thresholds on the installed PV power (i.e. 25\%, 50\% and 75\%). A block rate structure assumes a certain energy price for a given power level, in this study we assume three prices ranging from 2.395 CHF/kWh for very high power levels (i.e. higher than 75\% for the installed PV power), to 9.5 CHF/kWh for power levels below 25\% concerning the installed PV power. This block rate is different from the one used by Holweger et al. \cite{holweger2020mitigating}, which is applied on the import side. 
    
\end{enumerate}

\begin{table}[]
\centering
\scalebox{0.6}{
\begin{threeparttable}
\begin{tabular}{llllcccc}
\midrule
Tariff ID & Tariff   & \multicolumn{1}{c}{Season\tnote{a}} & Peak     &  \multicolumn{1}{p{4.5cm}}{\centering FiT \\{[}CHF/kWh{]}} & \multicolumn{1}{p{4.5cm}}{\centering  Import power cost  {[}CHF/kW{]}  } &\multicolumn{1}{p{4.5cm}}{\centering  Export power cost  {[}CHF/kW{]} } & \multicolumn{1}{p{2.4cm}}{\centering  Curtailment  {[}\%{]}}\\\midrule
0 & Reference DT                                     & -                                           & -        & 9.5       & 0                           & 0                           & -           \\\midrule
\multicolumn{1}{l}{\multirow{3}{*}{1}} & \multicolumn{1}{l}{\multirow{3}{*}{Variable DT}} & \multicolumn{1}{c}{\multirow{2}{*}{Summer}} & Off-peak (9-14h) & 5         & 0                           & 0                           &-          \\
  & \multicolumn{1}{l}{}                             & \multicolumn{1}{c}{}                        & Peak     & 9.5       & 0                           & 0                           & -         \\
  & \multicolumn{1}{l}{}                             & Winter                                      & -        & 9.5       & 0                           & 0                           & -        \\\midrule
2 & IRR Monthly                                      & -                                           & -        & 3.5-14.05 & 0                           & 0                           &-          \\\midrule
\multicolumn{1}{l}{\multirow{3}{*}{3}} & Curtailment 30                                   & -                                           & -        & 9.5       & 0                           & 0                           & 30\%        \\
 & Curtailment 50                                   & -                                           & -        & 9.5       & 0                           & 0                           & 50\%        \\
 & Curtailment 70                                   & -                                           & -        & 9.5       & 0                           & 0                           & 70\%        \\\midrule
4 & CT export daily                                  & -                                           & -        & 9.5       & 0                           & {[}0.8116, 1.0108, 1.2400{]} & -        \\\midrule
\multicolumn{1}{l}{\multirow{3}{*}{5}} & CT monthly 30                                    & -                                           & -        & 9.5       & 0.70                         & 0                           & 30\%        \\
  & CT monthly 50                                    & -                                           & -        & 9.5       & 0.70                         & 0                           & 50\%        \\
  & CT monthly 70                                    & -                                           & -        & 9.5       & 0.70                         & 0                           & 70\%        \\\midrule
\multicolumn{1}{l}{\multirow{3}{*}{6}} & CT daily 30                                      & -                                           & -        & 9.5       & {[}0.040, 0.070, 0.110{]} & 0                           & 30\%        \\
 & CT daily 50                                      & -                                           & -        & 9.5       & {[}0.040, 0.070, 0.110{]} & 0                           & 50\%        \\
 & CT daily 70                                     & -                                           & -        & 9.5       & {[}0.040, 0.070, 0.110{]} & 0                           & 70\%       \\\midrule
\multicolumn{1}{l}{\multirow{2}{*}{7}} & \multicolumn{1}{l}{\multirow{2}{*}{Block rate}}  & Summer                                      & -        & [9.5, 4.79, 2.395]\tnote{b} & 0 & 0                           & -      \\
  &                                                  & Winter                                      & -        & 9.5       & 0                           & 0                           & -      \\
\midrule
\end{tabular}
\begin{tablenotes}
\normalsize{\item[a] When the year is divided in two seasons, the summer runs from April to the end of August.}
\normalsize{\item[b] The amount paid to the client depends on the injected power. For instance, injection at very high power levels (i.e. higher than 75\% with respect to the installed PV power) is paid at 2.395 CHF/kWh.}
\end{tablenotes}

\end{threeparttable}}
\caption{Electricity tariff components depending on the bill structure used in this study. A single tariff at the import is considered, Reference DT with a total cost of 21.95 CHF/kWh on peak time (i.e. Monday to Friday from 6h until 22h) and 14.05 CHF/kWh at off-peak. }
\label{tab:tariffs}
\end{table}

\subsection{Resource sharing}\label{lab:mutua}
To assess the impact of the mutualization of assets, we consider the virtual fusion of the largest consumer and the largest producer to maximize the use of local PV electricity. This fusion is virtually made through the addition of the demand of both clients before the optimization. After the optimization, the remaining demand of the largest consumer is computed, accounting for the PV self-consumption, while the remaining demand is assigned to the highest producer. Then, the power flow analysis is performed.

\subsection{Modelling}
The three-step process involves a preliminary evaluation of demand allocation across each LV network, followed by consumer optimization that prioritizes economic reasoning, and concludes with an extensive analysis of the power flow within each respective network.
\subsubsection{Demand allocation}\label{sec:dem}
Holweger \textit{et al.} \cite{holweger2021privacy} introduced an approach to demand allocation using a two-stage Mixed-integer linear programming (MILP) optimization. The process consists of three phases: (i) estimating the annual energy consumption of each building in predefined zones using the RegBL data and SIA norms that describe building load categories \cite{holweger2021privacy, Flexi1_2015, Flexi2_2017}; (ii) allocating and scaling load curves to buildings based on their category and annual consumption through the minimization of the difference between the annual energy consumption of each building and reference load curves derived from real smart-meter data; and (iii) minimizing the difference between the sum of all building load curves and the measured transformer load curve in the second stage of optimization, while minimizing changes to each building load curve.

\subsubsection{Design and battery scheduling optimization}
Building upon previous work \cite{bloch2019impact, holweger2020mitigating}, the PV and battery sizing and operation for each building are optimized, minimizing the total cost of ownership, i.e., the sum of the annualized investment, maintenance, and operational cost (see Eq. \ref{eq:obj}), subject to power balance constraints (see Refs. \cite{bloch2019impact, holweger2020mitigating}, Supplementary Information Section 2 for further information on modelling, and Supplementary Information Section 3 for the model evaluation and validation). In this work, battery charging from the grid and battery export are forbidden, according to existing Swiss regulations for behind-the-meter storage systems. Therefore, the battery can uniquely be used to charge from the PV system and to cover the building's electricity needs. Perfect forecasting is used for both PV generation and electrical demand. 

\begin{equation}
\label{eq:obj}
minimize ~ (\textsc{capex}(P_{CAP}^{PV},E_{CAP}^{BAT}) + \textsc{opex}(P_{t}^{IMP},P_{t}^{EXP}))
\end{equation}

The objective function involves the installed PV capacity ($P_{CAP}^{PV}$) and battery size $(E_{CAP}^{BAT})$ in the cost calculation for capital expenditure (\textsc{capex}), whereas grid exchange power ($P_{t}^{IMP}$,$P_{t}^{EXP}$) is included in the cost calculation for operational expenditure (\textsc{opex}). Battery constraints and power balance constraints for the load, the grid, and the PV system are considered when optimizing the objective function. The operational costs only account for the cost of exchanging energy with the grid, excluding maintenance costs. Two types of prices, volumetric and capacity-based tariffs, are taken into account (refer to Eqs. \ref{eq:ge_vol} and \ref{eq:ge_cap}).

\begin{equation}
    \label{eq:OPEX}
     \textsc{OPEX} = \textsc{ox}_\text{ge}^\text{vol} + \textsc{ox}_\text{ge}^\text{pow}\\
\end{equation}
\begin{subequations}
	\begin{align}
	\label{eq:ge_vol}
	& \text{Volumetric tariff} & \textsc{ox}_\text{ge}^\text{vol} &= \sum_{t=1}^T \left[ P^\textsc{imp}_t \cdot  t^\textsc{imp}_t  - P^\textsc{exp}_t \cdot  t^\textsc{exp}_t\right ]\cdot \textsc{ts}_t\\
	\label{eq:ge_cap}
	& \text{Capacity-based tariff} & \textsc{ox}_\text{ge}^\text{pow} &= \sum_{k=1}^K  P^\textsc{max}_{k} \cdot  t^\textsc{max}_k
\end{align}
\end{subequations}

where $t^\textsc{imp}_t$ and $t^\textsc{exp}_t$ are the volumetric import and export tariff, respectively (in \,€/kWh), and $\textsc{ts}_t$ is the simulation time-step. For the capacity-based tariff, the maximum power of the billing period k, $P^\textsc{max}_k$ and $t^\textsc{max}_k$, is the maximum import (or export) power per billing period k. Two of the tariffs assessed in this study include a capacity-based component at the import (namely tariffs CT monthly and CT daily), and one includes a capacity-based component at the export (CT daily export).

To ensure a fair comparison between the alternative tariffs and the reference DT tariff, all tariffs must recover a similar share of grid costs across the three LV studied. Without this alignment, differences in results could arise due to the inherent structure of the tariffs rather than their performance or impact. To achieve this, the calibration of CT daily export, and CT monthly/daily curtailment tariffs was performed using an iterative, heuristic method. Specifically, the respective parameters shown in Table \ref{tab:tariffs} were adjusted to bring the total grid cost recovery close to that of the reference DT tariff, as reflected in Table \ref{tab:gridcost}, ensuring comparability in the results. However, it was not possible to achieve a perfect 100\% match for the different levels of curtailment with CT tariffs, as reflected by the slight variations in cost recovery. For example, CT monthly 70 and 30 tariffs recover 95\% and 110\% of grid costs, respectively, while curtailment-based tariffs recover between 89\% and 99\%. This difference arises due to the inherent difficulty in perfectly aligning curtailment-based tariffs with the total grid cost recovery target while maintaining the same network behavior. That discrepancy would not occur if each curtailment percentage had different pricing; however, this is not feasible, as the same pricing structure is applied to each three curtailment levels.

\subsubsection{Power flow analysis}
After obtaining the optimal schedule for the PV-coupled battery system and with the LV network's characteristics, we define a network graph using the Python packages Networkx and \textsc{pandapower}. We use the per-unit (p.u.) system for voltage and current in every node and line of the network, respectively.

To measure grid congestion, we use line loading level, which is calculated as the ratio between the current and the maximum nominal current of the line. We consider a line overloaded when its line loading level is above 100\%. Additionally, we distinguish between situations where the bus voltage at an injection point is above or below 1 p.u. and use the $95^{th}$ percentile of the bus voltage deviation. This helps us identify whether there is a local excess or deficit of energy. Voltage violations are defined as values above the (rather soft) limit of 1.1 p.u., and below 0.9 p.u. \cite{hartvigsson2021estimating}.

Lastly, we use load duration curves to evaluate the violations of the transformer power capacity for reverse power flow from LV to upstream medium-voltage (MV) networks, as well as the transformer permissible overload for different periods.

\subsection{Key performance indicators}
We present the results of the total amount of PV and storage installed per grid network, the amount of energy imported and exported, the maximum power feed-in and drawn for each zone, the total amount of curtailment within each zone, and the PV penetration (see Equation \ref{eq:pv_penetration}). Then, we analyze the PV load duration curve of the LV networks and the grid congestion. Additionally, we compare the grid cost recovery for each tariff explored. Finally, we present the levelized cost of electricity provision (LCOE), including PV and storage cost (see Equation \ref{eq:lcoe}), as well as the total cost per year (see Equation \ref{eq:total_cost}), the yearly profit, and the internal rate of return (see Equation \ref{eq:irr}).
\begin{align}
    \label{eq:pv_penetration}
    \text{PV Penetration} &= \frac{P_{\text{PV}}}{P_{\text{load}}} \times 100
\end{align}
where $P_\textsc{pv}$ and $P_\textsc{load}$ are the PV generation and consumption, respectively (in \,MWh).
\begin{align}
    \label{eq:lcoe}
    \text{LCOE} &= \frac{\sum_{t=1}^{T} \frac{C_t}{(1+r)^t}}{\sum_{t=1}^{T} \frac{E_t}{(1+r)^t}}
\end{align}
where $C_t$ is the total cost of investment plus operation and maintenance incurred in year $t$, $E_t$ is the amount of electricity produced in year $t$ (in MWh), $r$ is the discount rate, reflecting the time value of money, and $T$ is the total lifespan of the project (in years).
\begin{align}
    \label{eq:total_cost}
    \text{Total Cost} &= C_{\text{power}} + C_{\text{energy}}
\end{align}
where $C_\textsc{power}$ represents the costs associated with the power aspect, which includes fixed charges and capacity payments, and $C_\textsc{energy}$ being the costs incurred based on the actual energy consumed or produced, both in CHF/year.
\begin{align}
    \label{eq:irr}
    0 &= \sum_{t=0}^{T} \frac{R_t - C_t}{(1 + IRR)^t}
\end{align}
where $R_t$ represents the net cash inflow (revenue) in year $t$, $C_t$ is the total cash outflow (cost) in year $t$, and $T$ is the total project duration (in years). The internal rate of return (IRR) is the discount rate that makes the net present value of the cash flows equal to zero.

\section{Results}
The results are presented in three sections, covering the impact of tariffs on private investment decisions (i.e., PV and battery investment); the impact on the grid, covering the load duration curve at the transformer level as well as voltage deviation and line loading levels and grid usage; finally, the impact of shared resources on the grid is presented. It is important to highlight that in the various curtailment scenarios presented (30\%, 50\%, and 70\%), the percentage given represents the highest amount of power that can be fed into the grid concerning the nominal capacity of the PV system.

\subsection{Private PV and storage investment}
We compare the electricity tariffs first in terms of PV and storage capacity installed, as well as differences in energy import and export. Table \ref{tab:res} details these aspects for the optimized scenarios, including PV penetration, battery capacity installed, and yearly energy imports and exports from the building perspective.

Tariffs with a capacity-based component but without curtailment, such as CT export daily, severely impact PV penetration. For example, in the rural network, the CT export daily tariff results in only 5\% PV penetration, 95\% lower than the 94\% of the DT reference. This issue is also noted, though to a lesser extent, with the export tariff IRR monthly (18\%), Curtailment 30 (65\%), and DT variable (71\%). 

In contrast, tariffs such as Curtailment 50 and 70, as well as mixed tariffs with capacity-based components at the import and curtailment, allow for a high degree of PV installations to be operated within the three studied zones, all of which have values approaching 100\% PV penetration.
\label{sec:subsec3.1}
\begin{table}[H]
\centering
\scalebox{0.85}{
\begin{tabular}{llcccc}
  \hline
Network & Name & \multicolumn{1}{p{4cm}}{\centering PV penetration \\ {[}\%{]}} & \multicolumn{1}{p{2cm}}{\centering Battery Capacity \\ {[}kWh{]}} & \multicolumn{1}{p{2cm}}{\centering Energy Import \\ {[}MWh{]}} & \multicolumn{1}{p{2cm}}{\centering Energy Export  \\ {[}MWh{]}}\\ 
  \hline
Rural & DT reference & 94 & 1 & 107 & 1006 \\ 
  Rural & DT variable & 71 & 8 & 120 & 758 \\ 
  Rural & CT export daily & 5 & 22 & 167 & 17 \\ 
  Rural & Curtailment 70 & 94 & 4 & 106 & 1002 \\ 
  Rural & Curtailment 50 & 92 & 33 & 99 & 926 \\ 
  Rural & Curtailment 30 & 65 & 45 & 117 & 511 \\ 
  Rural & IRR monthly & 18 & 12 & 146 & 170 \\ 
  Rural & CT monthly 70 & 99 & 320 & 45 & 992 \\ 
  Rural & CT monthly 50 & 99 & 337 & 43 & 958 \\ 
  Rural & CT monthly 30 & 91 & 331 & 48 & 701 \\ 
  Rural & CT daily 70 & 98 & 187 & 61 & 997 \\ 
  Rural & CT daily 50 & 98 & 214 & 57 & 954 \\ 
  Rural & CT daily 30 & 88 & 221 & 63 & 681 \\ \hline
  Semi-urban & DT reference & 92 & 15 & 517 & 1271 \\ 
  Semi-urban & DT variable & 72 & 44 & 538 & 994 \\ 
  Semi-urban & CT export daily & 11 & 109 & 632 & 80 \\ 
  Semi-urban & Curtailment 70 & 92 & 29 & 513 & 1263 \\ 
  Semi-urban & Curtailment 50 & 92 & 123 & 488 & 1160 \\ 
  Semi-urban & Curtailment 30 & 66 & 185 & 523 & 622 \\ 
  Semi-urban & IRR monthly & 33 & 35 & 595 & 462 \\ 
  Semi-urban & CT monthly 70 & 93 & 897 & 342 & 1100 \\ 
  Semi-urban & CT monthly 50 & 93 & 944 & 335 & 1074 \\ 
  Semi-urban & CT monthly 30 & 92 & 1058 & 320 & 838 \\ 
  Semi-urban & CT daily 70 & 93 & 721 & 364 & 1122 \\ 
  Semi-urban & CT daily 50 & 93 & 778 & 355 & 1090 \\ 
  Semi-urban & CT daily 30 & 91 & 902 & 336 & 842 \\ \hline
  Urban & DT reference & 95 & 79 & 1110 & 1291 \\ 
  Urban & DT variable & 90 & 164 & 1090 & 1177 \\ 
  Urban & CT export daily & 40 & 413 & 1152 & 235 \\ 
  Urban & Curtailment 70 & 96 & 88 & 1106 & 1288 \\ 
  Urban & Curtailment 50 & 95 & 197 & 1072 & 1202 \\ 
  Urban & Curtailment 30 & 89 & 376 & 1038 & 870 \\ 
  Urban & IRR monthly & 65 & 179 & 1129 & 771 \\ 
  Urban & CT monthly 70 & 99 & 2574 & 744 & 975 \\ 
  Urban & CT monthly 50 & 99 & 2580 & 744 & 960 \\ 
  Urban & CT monthly 30 & 98 & 2686 & 731 & 828 \\ 
  Urban & CT daily 70 & 99 & 1583 & 806 & 1031 \\ 
  Urban & CT daily 50 & 99 & 1606 & 802 & 1018 \\ 
  Urban & CT daily 30 & 98 & 1771 & 780 & 852 \\ 
   \hline
\end{tabular}
}
\caption{Installed battery capacities and energy imports and exports per low-voltage network and tariff, for the PV penetration obtained from the optimization.}
\label{tab:res}
\end{table}

Regarding energy storage, tariffs with curtailment alone show a moderate effect on storage capacity installed, with average storage capacities of 34 kWh in rural areas, 112.33 kWh in semi-urban areas, and 220.33 kWh in urban areas. However, when capacity-based tariffs are combined with curtailment, storage use significantly increases. In rural areas, for example, the average installed storage increases by 7.9 times, from 34 kWh to 268.33 kWh. In semi-urban areas, storage capacity increases by 7.86 times, from 112.33 kWh to 883.33 kWh. The most pronounced increase is in urban areas, where the installed storage capacity grows by 9.68 times, from 220.33 kWh to 2133.33 kWh. This demonstrates the substantial boost in storage utilization when capacity-based tariffs are introduced alongside curtailment, with increases ranging from 2.5 to almost 10 times more storage compared to curtailment alone.

In terms of energy import and export, capacity-based tariffs at the import combined with curtailment significantly reduce energy imports, with reductions exceeding 59\% in some cases (e.g., CT monthly 50 in rural areas). In semi-urban and urban areas, the reductions in energy imports due to capacity-based tariffs range from 32\% to 35\%. Conversely, tariffs involving curtailment alone do not substantially impact energy imports. Regarding energy exports, tariffs with capacity-based components and curtailment can reduce exports by up to 36\% in urban areas (e.g., CT monthly 70). On the other hand, tariffs that involve curtailment alone can reduce exports by up to 51\% (e.g., Curtailment 30 in semi-urban areas), demonstrating a more significant impact on energy export reductions than capacity-based tariffs.

\subsection{Grid impact}

The results are presented for 100\% PV penetration. For further reference, Table \ref{tab:res} is replicated as Table \ref{tab:resMax} in the Annex, providing the same results but for the 100\% PV scenario. This scenario is particularly important as it represents a worst-case situation from the grid's perspective, pushing the limits of solar integration and highlighting the challenges for grid stability. This analysis offers valuable insights into their effectiveness in facilitating the Swiss energy transition and achieving national renewable energy goals.

Table \ref{tab:power} presents the maximum feed-in power, the maximum power drawn from the grid at the transformer station, and the total amount of PV curtailed (aggregated at the LV network level). Three key conclusions can be drawn from these results.

First, the level of curtailment plays a decisive role in setting the maximum feed-in power.  Across tariffs that impose the same curtailment level (e.g., CT daily 30, CT monthly 30, and Curtailment 30), the maximum feed-in power remains consistent or relatively close: 333 MW in rural areas, 439 MW in semi-urban areas, and 526 MW in urban areas. However, in the case of the CT export daily tariff, the maximum feed-in power is the lowest in the rural area, at 314 MW (or 33.6\% of the reference), and in the semi-urban area, at 324 MW (26.1\% of the reference). In contrast, the urban area has a much higher maximum feed-in power of 952 MW, or 69.7\% of the reference. This indicates that the tariff’s capacity-based component for exports may further restrict feed-in power in smaller or less populated networks. However, as seen above, this type of tariff heavily penalizes PV adoption and therefore should not be implemented in the early phases of PV development.

Second, the combination of curtailment with a capacity-based component at the import not only reduces the maximum feed-in power compared to the reference tariff but also decreases the maximum power drawn from the grid. This reduction in the maximum power drawn is notable in different tariffs depending on the location. For example, in the rural area, the CT daily 30 tariff draws 70 MW, which is 86.1\% of the reference value, compared to 71 MW (87.1\%) under the CT monthly 30 tariff. Similarly, in the semi-urban area, the CT daily 30 tariff draws 253 MW (84\%) of the reference, compared to 260 MW (86.4\%) for the CT monthly 30 tariff. In contrast, for the urban area, the CT monthly 30 tariff shows a greater reduction, with a maximum draw of 399 MW, which is 93.2\% of the reference, compared to 424 MW (98.9\%) under the CT daily 30 tariff.

Finally, regarding PV electricity curtailment, the maximum curtailment observed across the three LV networks was 75.06\% in the rural area under the CT export daily tariff. This indicates that, while the CT export daily tariff strongly limits feed-in power (314 MW, or 33.6\% of the reference), it leads to a significant amount of PV energy being curtailed (75.06\%), even though a consequent amount of storage is installed (141 kWh, as shown in Table \ref{tab:resMax}). Interestingly, for tariffs with a 50\% curtailment limit, less than 5\% of the total PV production had to be curtailed across the three LV networks. This demonstrates that a curtailment rate of 50\% does not necessarily result in excessive PV energy loss, possibly due to more efficient utilization of the remaining storage capacity.

\begin{table}[!htbp]
\scalebox{0.8}{
\begin{tabular}{llccccc}
\hline
Network    & Tariff name            & \multicolumn{1}{p{2cm}}{\centering Maximum Feed-in power \\ {[}MW{]}} & \multicolumn{1}{p{2cm}}{\centering Reference\\\%} & \multicolumn{1}{p{3cm}}{\centering Maximum Power drawn\\{[}MW{]}} &  \multicolumn{1}{p{2cm}}{\centering Reference\\\%} & \multicolumn{1}{p{2cm}}{\centering Curtailment\\\%}\\ \hline
Rural & CT daily 30     & 333  & \textbf{35.6\%} & 70  & \textbf{86.1\%} & 21.57\% \\ 
Rural & CT monthly 30   & 333  & \textbf{35.6\%} & 71  & 87.1\%          & 21.51\% \\
Rural & Curtailment 30  & 333  & \textbf{35.6\%} & 76  & 93.6\%          & 22.38\% \\
Rural & CT daily 50     & 551  & 58.9\%          & 72  & 88.5\%          & 4.30\%  \\
Rural & CT monthly 50   & 551  & 58.9\%          & 73  & 90.2\%          & 4.32\%  \\
Rural & Curtailment 50  & 551  & 58.9\%          & 80  & 97.7\%          & 4.83\%  \\
Rural & CT export daily & 314  & \textbf{33.6\%} & 71  & 86.9\%          & 75.06\% \\
Rural & CT daily 70     & 763  & 81.6\%          & 76  & 93.8\%          & 0.19\%  \\
Rural & CT monthly 70   & 763  & 81.6\%          & 80  & 97.8\%          & 0.21\%  \\
Rural & Curtailment 70  & 763  & 81.6\%          & 81  & 100.0\%         & 0.28\%  \\
Rural & Block rate      & 900  & 96.3\%          & 71  & 86.9\%          & 0.02\%  \\
Rural & DT variable     & 932  & 99.8\%          & 81  & 100.0\%         & 0\%     \\
Rural & IRR monthly     & 934  & 100.0\%         & 81  & 100.0\%         & 0\%     \\
Rural & DT reference    & 934  & 100.0\%         & 81  & 100.0\%         & 0\%     \\\hline
Semi-urban      & CT daily 30     & 439  & \textbf{35.3\%} & 253 & \textbf{84\%}   & 19.15\% \\
Semi-urban      & CT monthly 30   & 439  & \textbf{35.3\%} & 260 & 86.4\%          & 19.51\% \\
Semi-urban      & Curtailment 30  & 439  & \textbf{35.3\%} & 302 & 100.3\%         & 21.20\% \\
Semi-urban      & CT daily 50     & 729  & 58.6\%          & 260 & 86.3\%          & 3.39\%  \\
Semi-urban      & CT monthly 50   & 729  & 58.6\%          & 272 & 90.6\%          & 3.67\%  \\
Semi-urban      & Curtailment 50  & 729  & 58.6\%          & 301 & 100.0\%         & 4.66\%  \\
Semi-urban      & CT export daily & 324  & \textbf{26.1\%} & 302 & 100.5\%         & 68.35\% \\
Semi-urban      & CT daily 70     & 1015 & 81.6\%          & 280 & 93.1\%          & 0.09\%  \\
Semi-urban      & CT monthly 70   & 1015 & 81.6\%          & 287 & 95.5\%          & 0.13\%  \\
Semi-urban      & Curtailment 70  & 1015 & 81.6\%          & 301 & 99.9\%          & 0.25\%  \\
Semi-urban      & Block rate      & 1081 & 86.9\%          & 302 & 100.2\%         & 0\%     \\
Semi-urban      & IRR monthly     & 1244 & 100.0\%         & 301 & 100.1\%         & 0\%     \\
Semi-urban      & DT variable     & 1244 & 100.0\%         & 302 & 100.3\%         & 0\%     \\
Semi-urban      & DT reference    & 1244 & 100.0\%         & 301 & 100.0\%         & 0\%     \\\hline
Urban     & CT daily 30     & 526  & \textbf{38.6\%} & 424 & 98.9\%          & 11.05\%  \\
Urban     & CT monthly 30   & 526  & \textbf{38.6\%} & 399 & \textbf{93.2\%} & 11.52\%  \\
Urban     & Curtailment 30  & 526  & \textbf{38.6\%} & 432 & 100.7\%         & 13.53\% \\
Urban     & CT daily 50     & 873  & 64.0\%          & 424 & 99.0\%          & 1.58\%  \\
Urban     & CT monthly 50   & 873  & 64.0\%          & 399 & \textbf{93.2\%} & 1.57\%  \\
Urban     & Curtailment 50  & 873  & 64.0\%          & 431 & 100.7\%         & 2.26\%  \\
Urban     & CT export daily & 952  & 69.7\%          & 433 & 101.2\%         & 53.06\% \\
Urban     & Block rate      & 1163 & 85.3\%          & 433 & 101.2\%         & 0\%     \\
Urban     & CT daily 70     & 1199 & 87.8\%          & 424 & 98.9\%          & 0.05\%  \\
Urban     & CT monthly 70   & 1199 & 87.8\%          & 399 & \textbf{93.2\%} & 0.06\%  \\
Urban     & Curtailment 70  & 1199 & 87.8\%          & 428 & 100.0\%         & 0.12\%  \\
Urban     & IRR monthly     & 1365 & 100.0\%         & 434 & 101.4\%         & 0\%     \\
Urban     & DT variable     & 1365 & 100.0\%         & 429 & 100.2\%         & 0\%     \\
Urban     & DT reference    & 1365 & 100.0\%         & 428 & 100.0\%         & 0\%     \\ \hline

\end{tabular}
}
\caption{Maximum power drawn and injected at the transformer station per network and tariff. The best results with respect to the DT reference are in bold. All results presented are for a 100\% PV penetration (i.e. all the roofs are fully covered with PV).}
\label{tab:power}
\end{table}

\begin{figure}[H]
    \centering
    \includegraphics[scale=0.5]{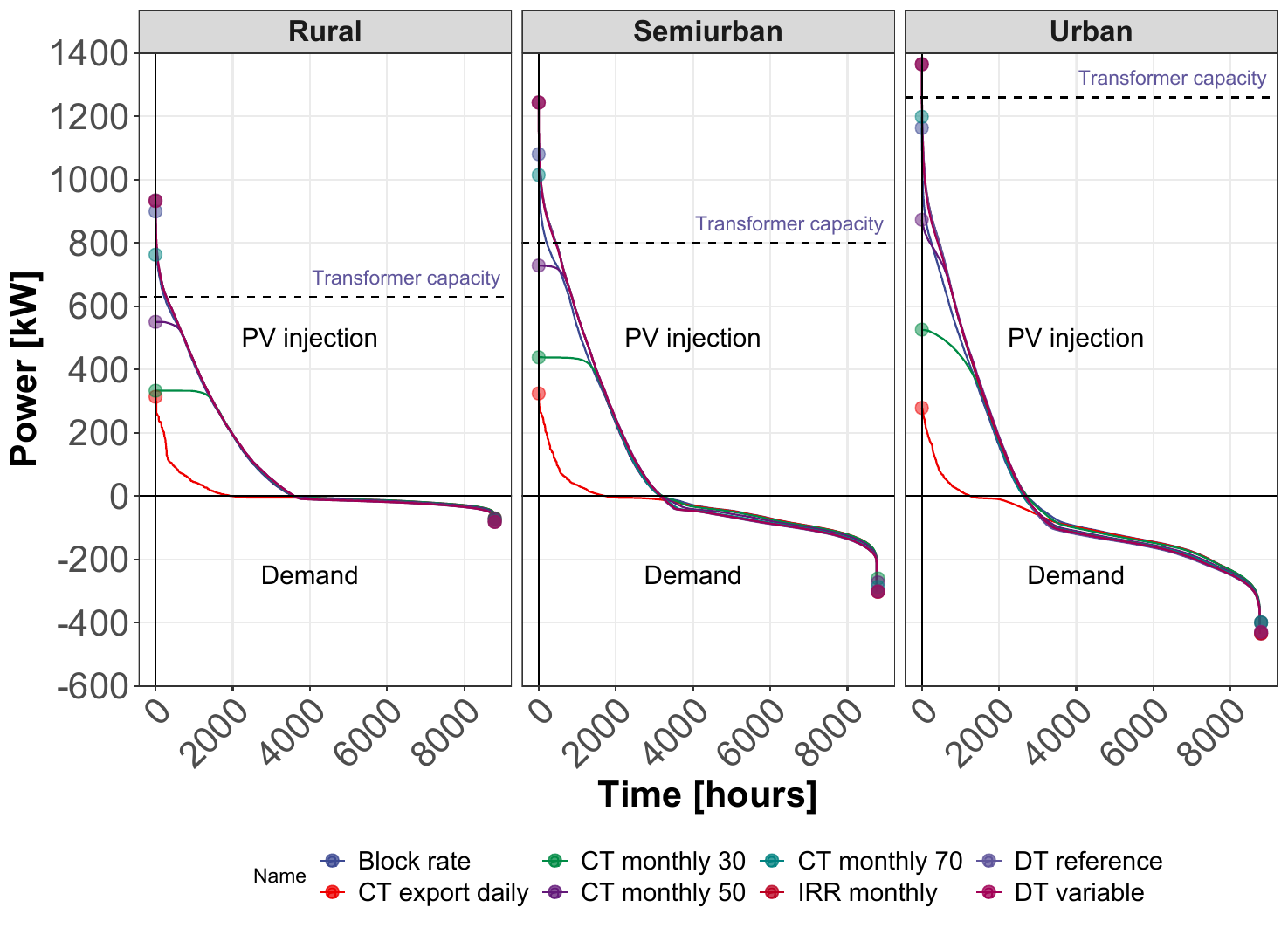}
    \caption{Load duration curve at the transformer. Dots on the vertical axis indicate the total installed PV capacity per scenario. Negative values indicate power flow from the high-voltage toward the low-voltage side}
    \label{fig:load}
\end{figure}

Since the maximum feed-in power is consistent across different tariffs with similar levels of curtailment, Fig \ref{fig:load} focuses on purely volumetric export prices, the block rate, and the three curtailment levels combined with a capacity-based component at the import monthly (i.e., CT monthly). The figure shows that only higher levels of curtailment (i.e., 50\% and 70\%) successfully prevent transformer overloading across the three types of networks. It is important to note that the tariffs IRR monthly, DT reference, and DT variable are grouped in the graph, as their values are nearly identical, making them visually indistinguishable.

\begin{figure}
    \centering
    \includegraphics[scale=0.45]{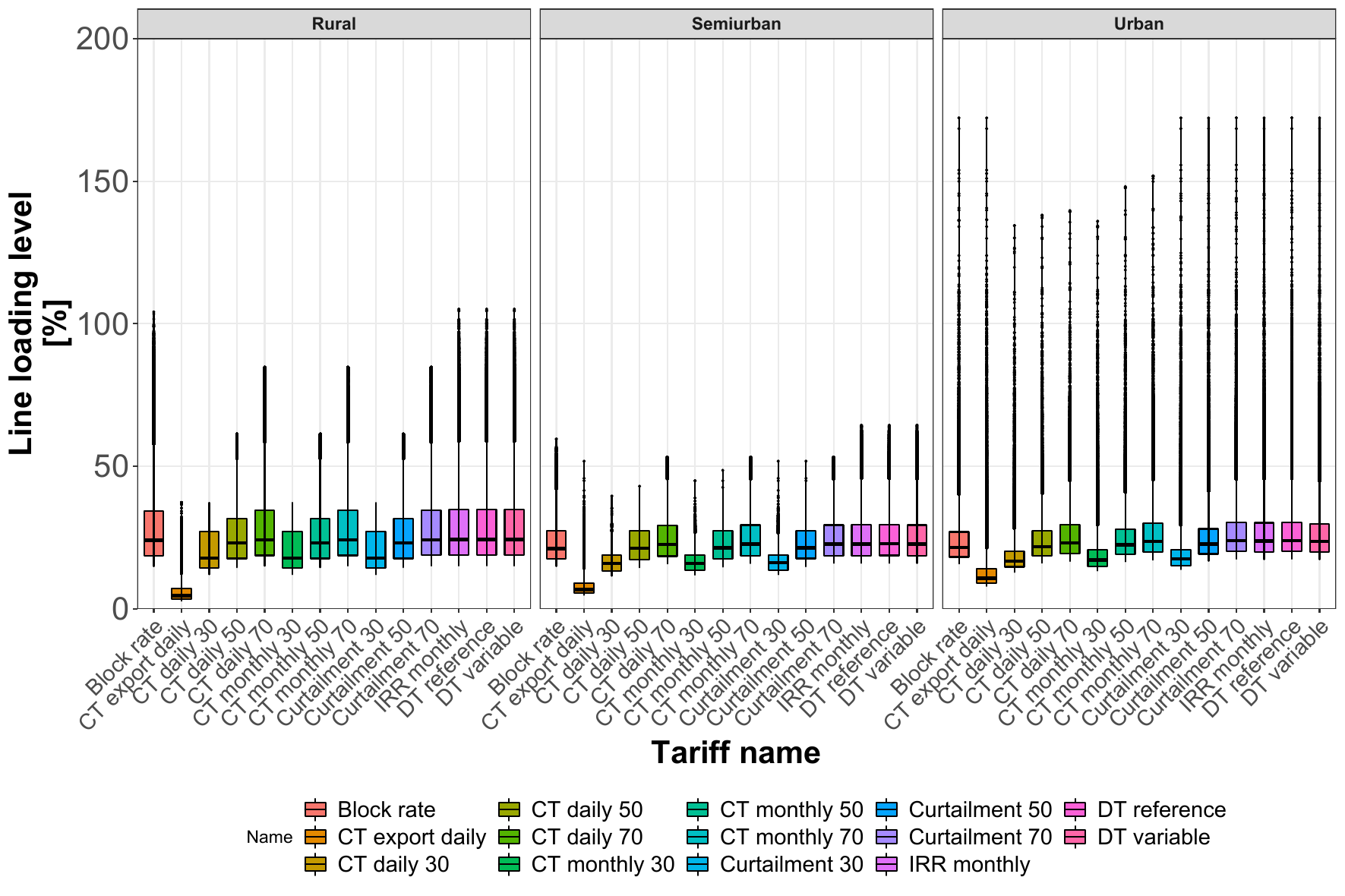}
    \caption{Line loading level distribution ($95^{th}$ percentile); Box plots show the median (horizontal line) and the IQR (box outline). The whiskers extend from the hinge to the highest and lowest values that are within 1.5 × IQR of the hinge, and the points represent the outliers.}
    \label{fig:lineload}
\end{figure}

The analysis of line loading at the node level, as shown in Figure \ref{fig:lineload}, reveals that tariffs with a capacity-based component can significantly reduce line overloading. CT export daily is the most effective, reducing line loading to 30\%–40\% in rural areas compared to the near-100\% levels of the reference tariff. The combination of curtailment and CT monthly tariffs further reduces loading to below 50\% networks. However, tariffs such as block rate, IRR monthly, DT reference, and DT variable occasionally exceed the 100\% line overloading threshold, highlighting their limitations in managing high-demand situations. This demonstrates the need for capacity-based or curtailment components to effectively manage line overloading, especially in networks facing more severe loading challenges.

In addition to the effectiveness observed in rural areas, semi-urban networks also benefit from capacity-based tariffs, albeit to a lesser extent. Tariffs like CT monthly 30 and Curtailment 50 maintain median line loadings between 40\% and 60\%, still far below the levels observed with the reference tariff. This suggests that the semi-urban grid benefits from capacity-based tariffs, experiencing less overloading pressure compared to the rural network.

However, it is worth highlighting that, in urban areas, the lines remain overloaded despite the use of different tariffs. The median line loading remains above 100\% even with tariffs such as CT monthly and varying levels of curtailment. This indicates that the capacity-based components are less effective in urban environments, where grid demands are higher. Further interventions or network reinforcements may be required to mitigate overloading in such areas.

\begin{figure}
    \centering
    \includegraphics[scale=0.5]{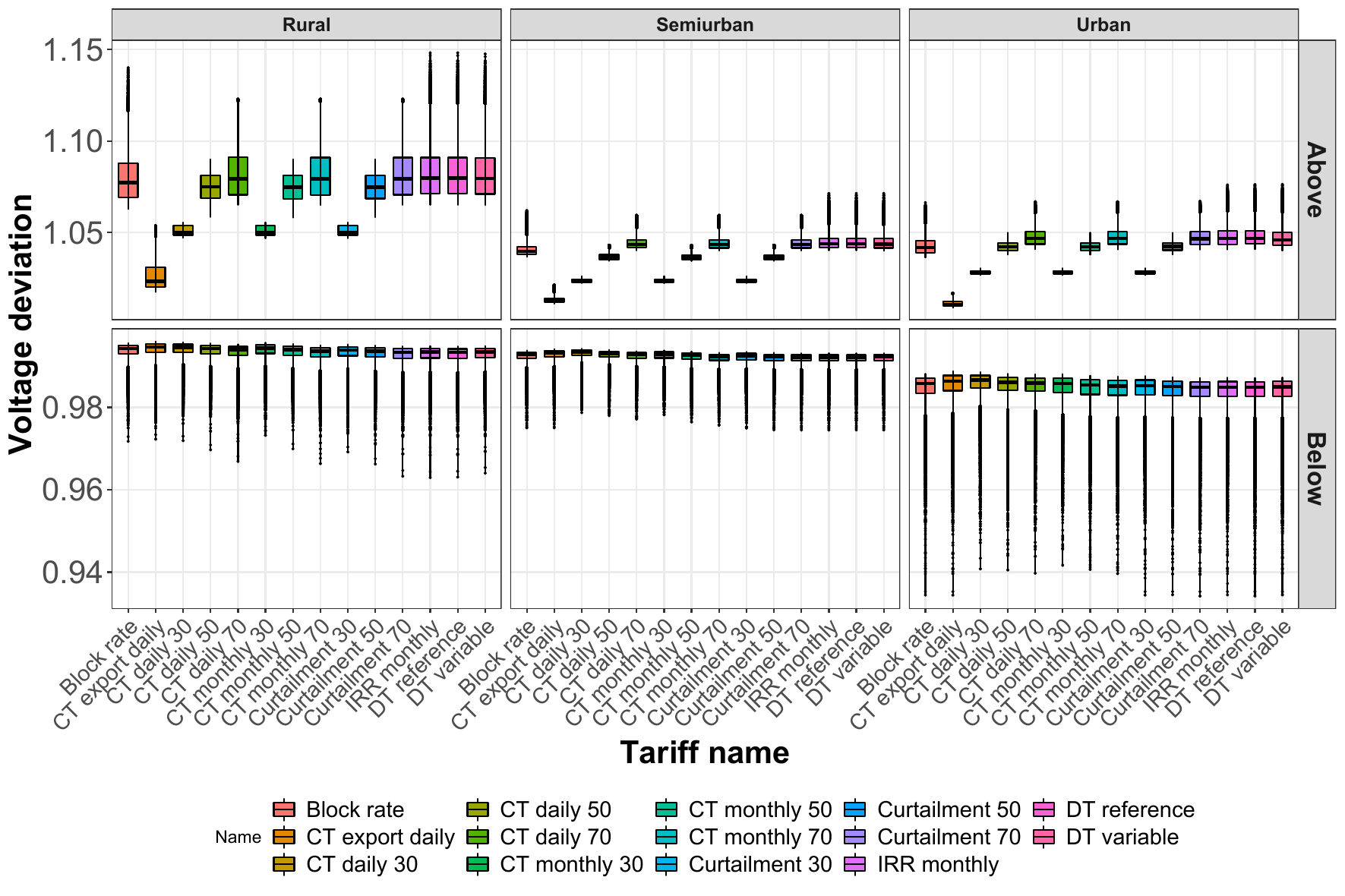}
    \caption{Voltage deviation distribution ($95^{th}$ percentile); Box plots show the median (horizontal line) and the IQR (box outline). The whiskers extend from the hinge to the highest and lowest values that are within 1.5 × IQR of the hinge, and the points represent the outliers.}
    \label{fig:volt}
\end{figure}

As for voltage deviation (Figure \ref{fig:volt}), positive deviations are particularly critical in the rural network, where aggressive curtailment can effectively reduce these deviations to stay within the limit of 1.1 p.u., according to European and Swiss regulations \cite{hartvigsson2021estimating}.
Under tariffs with 30 and 50 \% levels of curtailment, voltage deviations are reduced to values below 1.1 p.u. In contrast, without such curtailment, deviations frequently exceed the threshold in rural networks, highlighting the importance of curtailment for voltage regulation.

In semi-urban and urban networks, voltage deviations are generally not an issue. In these networks, deviations remain well within acceptable limits, with most tariffs resulting in deviations below 1.05 p.u. For instance, the CT monthly 50 and Curtailment 50 tariffs in urban areas result in deviations that are consistently below 1.05 p.u., suggesting that voltage deviation is not a pressing concern here. This indicates that while tariffs with a capacity-based component are effective in managing both voltage deviations and line loading, the specific need for curtailment varies depending on the network type. Rural networks require more aggressive curtailment to handle both voltage deviations and line overloading, while urban networks do not face significant voltage deviation issues but still struggle with line overloading.

\label{sec:subsec3.2}
\subsection{Economic results}
The economic results for scenarios with maximum PV penetration and optimized battery capacity, shown in Table \ref{tab:econResultsmax}, demonstrate significant variation across different tariff structures. All reported values represent averages of the studied buildings. 

The lowest Levelized Cost of Electricity (LCOE) is achieved with the DT reference tariff, reaching 0.13 CHF/kWh in urban areas, followed closely by the Curtailment 70 tariff with an LCOE of 0.13 CHF/kWh. In contrast, the CT export daily tariff shows a significantly higher LCOE of 0.74 CHF/kWh in rural zones, contributing to its poor economic performance. 

However, CT export daily and IRR monthly tariffs exhibit particularly negative economic outcomes, with some cases displaying negative internal rates of return (IRR) and financial losses. For example, the CT export daily tariff in rural areas has an IRR of -8.92\% and a loss of -37827 CHF annually, while the IRR monthly tariff in semi-urban areas results in an IRR of -0.99\% and a loss of -888 CHF annually. These tariffs are particularly detrimental, as they not only hinder PV adoption but also remain unfavorable even after 100\% PV penetration, leading to negative returns and financial disadvantages for users.

The losses and negative IRR associated with the CT export daily tariff across all zones can be attributed to its structure. While it offers a fixed export price, the significant capacity-based component on exports results in high penalties for injecting power into the grid. For instance, in urban areas, the CT export daily tariff results in a loss of -1063 CHF annually, despite having a relatively low (but still the highest within the urban area) LCOE of 0.24 CHF/kWh. This makes the tariff economically disadvantageous, especially for buildings with high PV generation but limited self-consumption, as they are penalized for exporting excess energy.

Conversely, tariffs with a capacity-based component at the import, such as CT daily and CT monthly, consistently deliver higher IRRs and annual profits. These tariffs allow for better optimization of the battery and PV systems, as their structure helps reduce grid dependency while avoiding the heavy export penalties seen with CT export daily. The CT daily tariff stands out as having the highest IRR and profits. For example, in urban areas, the CT daily 70 tariff achieves an IRR of 7.62\% and an annual profit of 36391 CHF, while the CT monthly 70 tariff produces an IRR of 7.38\% and a profit of 34193 CHF. It is followed closely by the CT monthly 70 tariff, which achieves an IRR of 7.38\% and a profit of 34,193 CHF in the same zone. Both tariffs demonstrate that capacity-based tariffs strike a favorable balance between cost savings and maximizing the value of energy produced and consumed locally.

\begin{table}[ht!]
\centering
\scalebox{0.8}{
\begin{tabular}{llcccc}
  \hline
Network & Name & \multicolumn{1}{p{2cm}}{\centering LCOE  \\ {[}CHF/kWh{]}}  & \multicolumn{1}{p{2cm}}{\centering IRR  \\ {[}\%{]}} & \multicolumn{1}{p{2cm}}{\centering Total cost  \\ {[}CHF p.a.{]}}  & \multicolumn{1}{p{2cm}}{\centering Profit  \\ {[}CHF p.a.{]}}  \\ 
  \hline
Rural & DT reference & 0.16 & 4.87 & 543 & 16104 \\ 
  Rural & DT variable & 0.26 & 3.17 & 526 & 6606 \\ 
  Rural & CT export daily & 0.74 & -8.92 & 748 & -37827 \\ 
  Rural & Curtailment 70 & 0.16 & 4.84 & 538 & 15947 \\ 
  Rural & Curtailment 50 & 0.20 & 4.32 & 480 & 13184 \\ 
  Rural & Curtailment 30 & 0.31 & 2.34 & 421 & 2321 \\ 
  Rural & IRR monthly & 0.39 & 0.74 & 529 & -5880 \\ 
  Rural & Block rate & 0.20 & 4.33 & 528 & 9568 \\ 
  Rural & CT monthly 70 & 0.18 & 5.12 & 563 & 17824 \\ 
  Rural & CT monthly 50 & 0.21 & 4.67 & 472 & 15426 \\ 
  Rural & CT monthly 30 & 0.32 & 2.79 & 400 & 4836 \\ 
  Rural & CT daily 70 & 0.18 & 5.22 & 528 & 18511 \\ 
  Rural & CT daily 50 & 0.20 & 4.77 & 463 & 16090 \\ 
  Rural & CT daily 30 & 0.32 & 2.90 & 403 & 5459
  \\ \hline
  Semi-urban & DT reference & 0.20 & 2.55 & 1246 & 12556 \\ 
  Semi-urban & DT variable & 0.24 & 1.10 & 1214 & 6767 \\ 
  Semi-urban & CT export daily & 0.43 & -6.17 & 1105 & -17684 \\ 
  Semi-urban & Curtailment 70 & 0.20 & 2.52 & 1233 & 12489 \\ 
  Semi-urban & Curtailment 50 & 0.22 & 3.31 & 1154 & 10851 \\ 
  Semi-urban & Curtailment 30 & 0.26 & 1.63 & 1068 & 4368 \\ 
  Semi-urban & IRR monthly & 0.30 & -0.99 & 1224 & -888 \\ 
  Semi-urban & Block rate & 0.21 & 2.26 & 1220 & 9840 \\ 
  Semi-urban & CT monthly 70 & 0.21 & 2.87 & 1286 & 14418 \\ 
  Semi-urban & CT monthly 50 & 0.23 & 3.74 & 1165 & 13238 \\ 
  Semi-urban & CT monthly 30 & 0.27 & 2.17 & 1064 & 7079 \\ 
  Semi-urban & CT daily 70 & 0.22 & 3.01 & 1268 & 15285 \\ 
  Semi-urban & CT daily 50 & 0.23 & 3.91 & 1168 & 14203 \\ 
  Semi-urban & CT daily 30 & 0.28 & 2.38 & 1077 & 8108 \\ \hline
  Urban & DT reference & 0.13 & 6.68 & 3005 & 28627 \\ 
  Urban & DT variable & 0.15 & 5.47 & 2906 & 22241 \\ 
  Urban & CT export daily & 0.24 & -0.27 & 2606 & -4063 \\ 
  Urban & Curtailment 70 & 0.13 & 6.66 & 2992 & 28621 \\ 
  Urban & Curtailment 50 & 0.14 & 6.35 & 2874 & 27566 \\ 
  Urban & Curtailment 30 & 0.16 & 5.12 & 2700 & 21732 \\ 
  Urban & IRR monthly & 0.18 & 3.82 & 2898 & 13825 \\ 
  Urban & Block rate & 0.14 & 6.32 & 2920 & 25305 \\ 
  Urban & CT monthly 70 & 0.15 & 7.38 & 3211 & 34193 \\ 
  Urban & CT monthly 50 & 0.16 & 7.15 & 3097 & 33648 \\ 
  Urban & CT monthly 30 & 0.18 & 6.06 & 2859 & 28692 \\ 
  Urban & CT daily 70 & 0.15 & 7.62 & 3141 & 36391 \\ 
  Urban & CT daily 50 & 0.16 & 7.42 & 3063 & 35892 \\ 
  Urban & CT daily 30 & 0.18 & 6.37 & 2861 & 31119 \\ \hline
\end{tabular}}
\caption{Economic results with the maximum PV capacity and an optimized battery design.}
\label{tab:econResultsmax}
\end{table}

Table \ref{tab:gridcost} presents the annual grid cost recovery per network based on the tariff applied. The DT reference tariff results in a total grid cost recovery of 168,844 CHF per year, serving as the baseline for comparison with 100\% reference grid cost recovery. The DT variable tariff slightly reduces the grid cost to 164,375 CHF, which is 97\% of the reference cost. The IRR monthly and block rate tariffs recover 164,621 CHF and 165,009 CHF, respectively, both representing 97\% and 98\% of the reference grid cost.

However, some tariffs show higher grid cost recovery because it was not possible to achieve a perfect 100\% match for the different levels of curtailment with CT tariffs, as reflected by the slight variations in cost recovery. For example, CT monthly 70 and CT monthly 30 tariffs recover 95\% and 110\% of grid costs, respectively, while curtailment-based tariffs recover between 89\% and 99\%. This discrepancy arises due to the inherent difficulty in perfectly aligning curtailment-based tariffs with the total grid cost recovery target while maintaining the same network behavior. If different pricing were applied to each curtailment level, this variation could be minimized, but this is not feasible as the same pricing structure is applied across all curtailment levels.

\begin{table}[H]
\centering
\scalebox{0.85}{
\begin{tabular}{lrrrrr}
  \hline
Name            & Rural & Semi-urban  & Urban & Total &\multicolumn{1}{p{2cm}}{\centering Reference\\\%}  \\ \hline
DT reference    & 9940       & 50420  & 108485    & 168844 & 100\% \\
DT variable     & 9671       & 49332  & 105373    & 164375 & 97\%  \\
CT export daily & 19526      & 51447 & 101009    & 171982 & 101\% \\
Curtailment 70  & 9847       & 49962  & 108088    & 167897 & 99\%  \\
Curtailment 50  & 8939       & 47228  & 104333    & 160501 & 95\%  \\
Curtailment 30  & 7946       & 44141  & 98764     & 150851 & 89\%  \\
IRR monthly     & 9723       & 49708  & 105190    & 164621 & 97\%  \\
Block rate      & 9692       & 49545  & 105772    & 165009 & 98\%  \\
CT monthly 70   & 8085      & 44971  & 107856    & 160912 & 95\% \\
CT monthly 50   & 9473      & 49155  & 116510    & 175139 & 104\% \\
CT monthly 30   & 11169      & 53990  & 120552    & 185711 & 110\% \\
CT daily 70     & 7978       & 46111  & 108173    & 162262 & 96\% \\
CT daily 50     & 9141       & 50066  & 116152    & 175359 & 104\% \\
CT daily 30     & 10420      & 54419  & 119229    & 184068 & 109\% \\\hline
      
\end{tabular}}
\caption{Grid cost recovery per network and tariff, in € per year, for the maximum amount of PV installed per network.}
\label{tab:gridcost}
\end{table}
\subsection{Resource sharing impact}

The main results of the impact of mutualization between the largest consumer and the largest producer within each LV network, are presented in Table \ref{tab:sharing}. Overall, sharing assets in the networks does not affect the penetration of PV. For all networks, the total battery capacity is less than the original battery capacity, with differences ranging from -8 kWh to -1 kWh. This means that sharing reduces the overall battery capacity in the networks. For most networks, shared energy import is slightly less than the original energy import, with differences ranging from -2 MWh to -1 MWh. However, for the semi-urban scenario, DT variable and Curtailment 50, shared energy import is the same as the original energy import. Finally, shared energy export is slightly lower than the original energy export, with differences ranging from -2 MWh to -1 MWh. However, for the rural scenario, Curtailment 50, shared energy export is the same as the original energy export.

\begin{table}[!ht]
    \centering
    \scalebox{0.45}{
    \begin{tabular}{llcccccccccccc}
    \hline
        Network & Name & \multicolumn{1}{p{2.5cm}}{\centering \textbf{Original}\\PV penetration \\ {[}\%{]}}  & \multicolumn{1}{p{2.5cm}}{\centering \textbf{Shared}\\PV penetration \\ {[}\%{]}}  & \multicolumn{1}{p{2cm}}{\centering Difference \\ {[}p.p.{]}} & \multicolumn{1}{p{2cm}}{\centering \textbf{Original}\\Battery Capacity \\ {[}kWh{]}} & \multicolumn{1}{p{2cm}}{\centering \textbf{Shared}\\Battery Capacity \\ {[}kWh{]}} & \multicolumn{1}{p{2cm}}{\centering Difference\\ {[}kWh{]}} & \multicolumn{1}{p{2cm}}{\centering \textbf{Original}\\Energy Import \\ {[}MWh{]}}  & \multicolumn{1}{p{2cm}}{\centering \textbf{Shared}\\Energy Import \\ {[}MWh{]}}  & \multicolumn{1}{p{2cm}}{\centering Difference\\ {[}MWh{]}} & \multicolumn{1}{p{2cm}}{\centering \textbf{Original}\\Energy Export \\ {[}MWh{]}} & \multicolumn{1}{p{2cm}}{\centering \textbf{Shared}\\Energy Export \\ {[}MWh{]}} & \multicolumn{1}{p{2cm}}{\centering Difference\\ {[}MWh{]}} \\ 
    \hline
        Rural & DT reference & 94 & 94 & 0 & 1 & 1 & 0 & 107 & 106 & -1 & 1006 & 1004 & -2 \\ 
        Rural & DT variable & 71 & 71 & 0 & 8 & 8 & 0 & 120 & 119 & -1 & 758 & 757 & -1 \\ 
        Rural & Curtailment 70 & 94 & 94 & 0 & 4 & 3 & -1 & 106 & 105 & -1 & 1002 & 1001 & -1 \\
        Rural & Curtailment 50 & 92 & 92 & 0 & 33 & 31 & -2 & 99 & 98 & -1 & 926 & 926 & 0 \\ 
        Rural & Curtailment 30 & 65 & 65 & 0 & 45 & 43 & -2 & 117 & 116 & -1 & 511 & 510 & -1 \\ 
    \hline
        Semi-urban & DT reference & 92 & 92 & 0 & 15 & 12 & -3 & 517 & 516 & -1 & 1271 & 1270 & -1 \\ 
        Semi-urban & DT variable & 72 & 72 & 0 & 44 & 39 & -5 & 538 & 537 & -1 & 994 & 993 & -1 \\ 
        Semi-urban & Curtailment 70 & 92 & 92 & 0 & 29 & 24 & -5 & 513 & 512 & -1 & 1263 & 1262 & -1 \\ 
        Semi-urban & Curtailment 50 & 92 & 92 & 0 & 123 & 115 & -8 & 488 & 488 & 0 & 1160 & 1161 & 1 \\
        Semi-urban & Curtailment 30 & 66 & 66 & 0 & 185 & 178 & -7 & 523 & 524 & 1 & 622 & 622 & 0 \\ 
    \hline
        Urban & DT reference & 95 & 95 & 0 & 79 & 76 & -3 & 1110 & 1108 & -2 & 1291 & 1290 & -1 \\ 
        Urban & DT variable & 90 & 90 & 0 & 164 & 160 & -4 & 1090 & 1089 & -1 & 1177 & 1176 & -1 \\ 
        Urban & Curtailment 70 & 96 & 96 & 0 & 88 & 85 & -3 & 1106 & 1105 & -1 & 1288 & 1287 & -1 \\ 
        Urban & Curtailment 50 & 95 & 95 & 0 & 197 & 194 & -3 & 1072 & 1071 & -1 & 1202 & 1201 & -1 \\
        Urban & Curtailment 30 & 89 & 89 & 0 & 376 & 371 & -5 & 1038 & 1037 & -1 & 870 & 870 & 0 \\ 
    \hline
    \end{tabular}}
    \caption{Comparison of the results from the optimization of shared assets and the original (individual) optimization. }
    \label{tab:sharing}
\end{table}

\begin{figure}[H]
    \centering
    \includegraphics[scale=0.4]{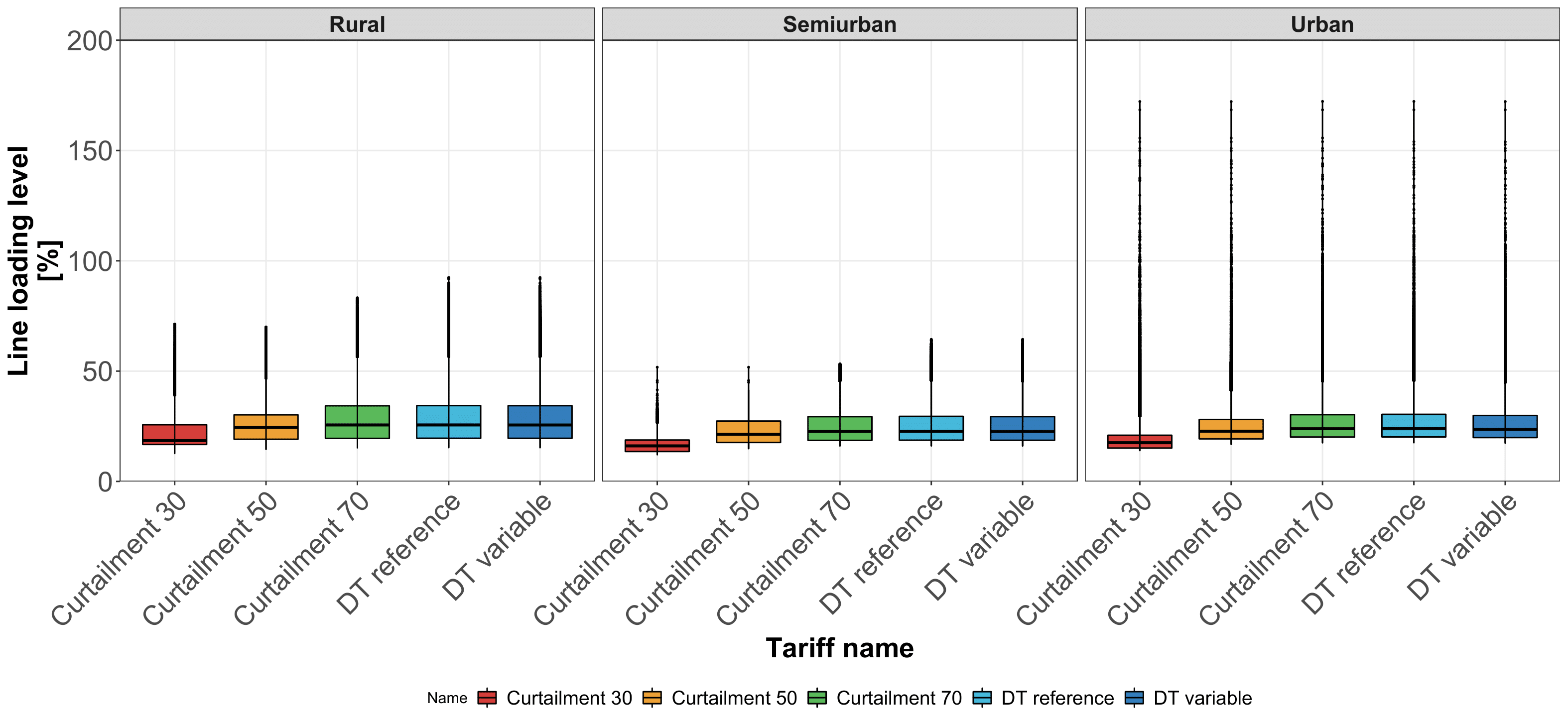}
    \caption{Line loading level distribution ($95^{th}$ percentile) under a resource sharing scenario with the maximum PV capacity and an optimized battery design; Box plots show the median (horizontal line) and the IQR (box outline). The whiskers extend from the hinge to the highest and lowest values that are within 1.5 × IQR of the hinge, and the points represent the outliers.}
    \label{fig:lineloadP2Pmax}
\end{figure}

For the maximum amount of PV in a scenario of shared assets, there is a reduction in the line loading level in the rural network as shown in Fig. \ref{fig:lineloadP2Pmax}, in particular, regarding the outliers that are now below the 100\% level (when compared with Fig. \ref{fig:lineload}). The maximum values for the rural scenario decreased in the cases of DT reference and DT variable (105\% to 92.4\% in both cases), whereas the median values increased from 24.3\% to 25.6\%. On the contrary, in the cases with high degrees of curtailment (i.e. Curtailment 30 and 50), the maximum loading level increase from 37.2\% to 71.3\% and from 61.2\% to 70\%. In the semi-urban and urban networks, no significant difference can be appreciated. As for the voltage deviations, these are higher in the rural network when compared with the results without shared assets (see Fig. \ref{fig:volt}). The median of the distribution in the rural scenario increased between 0.01-0.02 p.u. (e.g. from 1.05 to 1.07 for Curtailment 30). The results for the scenario where the PV and battery design are optimized are presented in Fig. \ref{fig:lineloadP2P} and Fig. \ref{fig:voltP2P}.

\begin{figure}[H]
    \centering
    \includegraphics[scale=0.4]{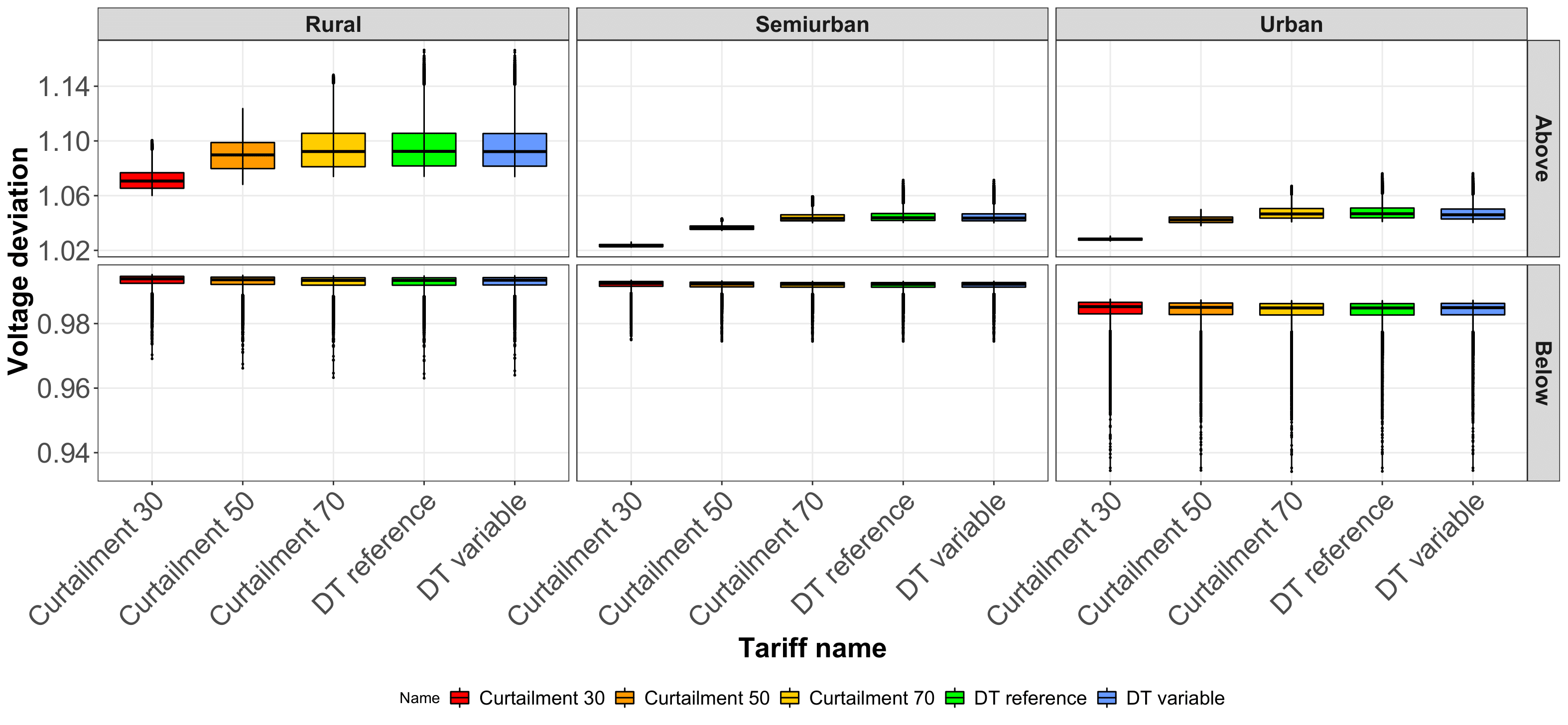}
    \caption{Voltage deviation distribution ($95^{th}$ percentile) under a resource sharing scenario with the maximum PV capacity and an optimized battery design; Box plots show the median (horizontal line) and the IQR (box outline). The whiskers extend from the hinge to the highest and lowest values that are within 1.5 × IQR of the hinge, and the points represent the outliers.}
    \label{fig:voltP2Pmax}
\end{figure}
\section{Conclusion}
The results of this study highlight the complexities and trade-offs involved in designing electricity tariffs that promote renewable energy integration while maintaining the stability of LV networks. One of the main findings is that the IRR monthly tariff, which is designed to incentivize winter production, may inadvertently penalize the installation of PV systems aimed at maximizing summer generation, thereby hindering the goal of increasing solar energy output. In contrast, the DT variable shows better economic performance in this regard.

Moreover, tariffs with a capacity-based component at the export, while effective in reducing grid overloading, significantly discourage PV installations due to the high financial penalties associated with power exports. Even after installations are completed, these tariffs can lead to economic losses, further complicating the adoption of solar energy solutions.

Curtailment proves to be a straightforward mechanism to manage the impacts of high PV penetration on LV networks, particularly for reducing transformer overloading and voltage violations in rural areas. However, the findings show that a curtailment level of 70\% provides only marginal benefits compared to a 50\% threshold. Curtailing at 50\% strikes a balance between grid stability and maintaining the economic feasibility of PV systems, although it may face resistance from the public due to concerns over lost renewable energy production.

Tariffs that combine a capacity-based component at the import with curtailment stand out as a promising approach. These tariffs not only alleviate stress on the grid by reducing peak demand but also represent an incentive for the integration of energy storage systems. Unlike pure curtailment mechanisms, capacity-based tariffs do not affect further the PV installation. This combination of curtailment and capacity-based tariffs provides a comprehensive strategy for addressing the future challenges posed by increased electrification of heating and transport systems.

In conclusion, the combination of capacity-based tariffs at the import and moderate curtailment (around 50\%) emerges as the most effective and economically viable strategy for managing PV integration in LV networks. This approach encourages the adoption of PV and storage systems while preventing grid overloads and maintaining voltage stability, positioning it as a practical solution for transitioning toward a more electrified and renewable energy future. While the block rate tariff is marginally better than purely volumetric tariffs, it fails to incentivize storage deployment and does not perform as well as those with significant curtailment. Therefore, capacity-based tariffs must be carefully calibrated to minimize economic disparities compared to reference tariffs, as the number of buildings significantly influences economic outcomes.

Finally, resource sharing between buildings shows promise, especially in rural areas, by reducing the need for individual battery capacity and reducing line overloads. Nevertheless, its impact on semi-urban and urban networks is limited and should be considered in tandem with other measures.

\bibliographystyle{vancouver}
\bibliography{biblio}
\appendix
\section{Results for full PV penetration}
\setcounter{table}{0}
\begin{table}[H]
\centering
\scalebox{0.8}{
\begin{tabular}{llcccc}
  \hline
network & Name & \multicolumn{1}{p{4cm}}{\centering PV penetration \\ {[}\%{]}} & \multicolumn{1}{p{2cm}}{\centering Battery Capacity \\ {[}kWh{]}} & \multicolumn{1}{p{2cm}}{\centering Energy Import \\ {[}MWh{]}} & \multicolumn{1}{p{2cm}}{\centering Energy Export  \\ {[}MWh{]}}\\ 
  \hline
Rural & DT reference    & 100   & 1            & 101           & 1069          \\
Rural & DT variable     & 100   & 10           & 98            & 1066          \\
Rural & CT export daily & 100   & 141          & 66            & 152           \\
Rural & Curtailment 70  & 100   & 4            & 100           & 1065          \\
Rural & Curtailment 50  & 100   & 37           & 91            & 1001          \\
Rural & Curtailment 30  & 100   & 78           & 81            & 785           \\
Rural & IRR monthly     & 100   & 9            & 99            & 1066          \\
Rural & Block rate      & 100   & 9            & 98            & 1066          \\
Rural & CT monthly 70   & 100   & 37          & 91            & 1056           \\
Rural & CT monthly 50   & 100   & 90          & 78            & 994           \\
Rural & CT monthly 30   & 100   & 139          & 67            & 781          \\
Rural & CT daily 70     & 100   & 54          & 86            & 1051           \\
Rural & CT daily 50     & 100   & 93          & 77            & 993           \\
Rural & CT daily 30     & 100   & 134          & 68            & 781          \\\hline
Semi-urban      & DT reference    & 100   & 15  & 511           & 1388          \\
Semi-urban      & DT variable     & 100   & 50  & 500           & 1376          \\
Semi-urban      & CT export daily & 100   & 510          & 392           & 147          \\
Semi-urban      & Curtailment 70  & 100   & 29           & 507           & 1379          \\
Semi-urban      & Curtailment 50  & 100   & 129          & 479           & 1278          \\
Semi-urban      & Curtailment 30  & 100   & 258          & 448           & 975           \\
Semi-urban      & IRR monthly     & 100   & 41           & 504           & 1380          \\
Semi-urban      & Block rate      & 100   & 43           & 502           & 1378          \\
Semi-urban      & CT monthly 70   & 100   & 127         & 479           & 1352           \\
Semi-urban      & CT monthly 50   & 100   & 283          & 442           & 1255          \\
Semi-urban      & CT monthly 30   & 100   & 436          & 409           & 961          \\
Semi-urban      & CT daily 70     & 100   & 208          & 460           & 1332           \\
Semi-urban      & CT daily 50     & 100   & 337          & 430           & 1247          \\
Semi-urban      & CT daily 30     & 100   & 476 & 401           & 958          \\\hline
Urban     & DT reference    & 100   & 80  & 1100          & 1371          \\
Urban     & DT variable     & 100   & 173 & 1069          & 1338          \\
Urban     & CT export daily & 100   & 843          & 909           & 132           \\
Urban     & Curtailment 70  & 100   & 91           & 1096          & 1365          \\
Urban     & Curtailment 50  & 100   & 210          & 1058          & 1283          \\
Urban     & Curtailment 30  & 100   & 421          & 1002          & 1003          \\
Urban     & IRR monthly     & 100   & 191          & 1067          & 1336          \\
Urban     & Block rate      & 100   & 160          & 1073          & 1342          \\
Urban     & CT monthly 70   & 100   & 301         & 1032           & 1299           \\
Urban     & CT monthly 50   & 100   & 423         & 1001           & 1236           \\
Urban     & CT monthly 30   & 100   & 718         & 935           & 972           \\
Urban     & CT daily 70     & 100   & 459         & 993           & 1257           \\
Urban     & CT daily 50     & 100   & 546         & 972           & 1210          \\
Urban     & CT daily 30     & 100   & 800         & 919           & 964          \\ 
   \hline
\end{tabular}
}
\caption{Installed battery capacities and energy imports and exports per low-voltage network and per tariff, for the results with maximum PV.}
\label{tab:resMax}
\end{table}

\setcounter{figure}{0}
\begin{figure}[H]
    \centering
    \includegraphics[scale=0.35]{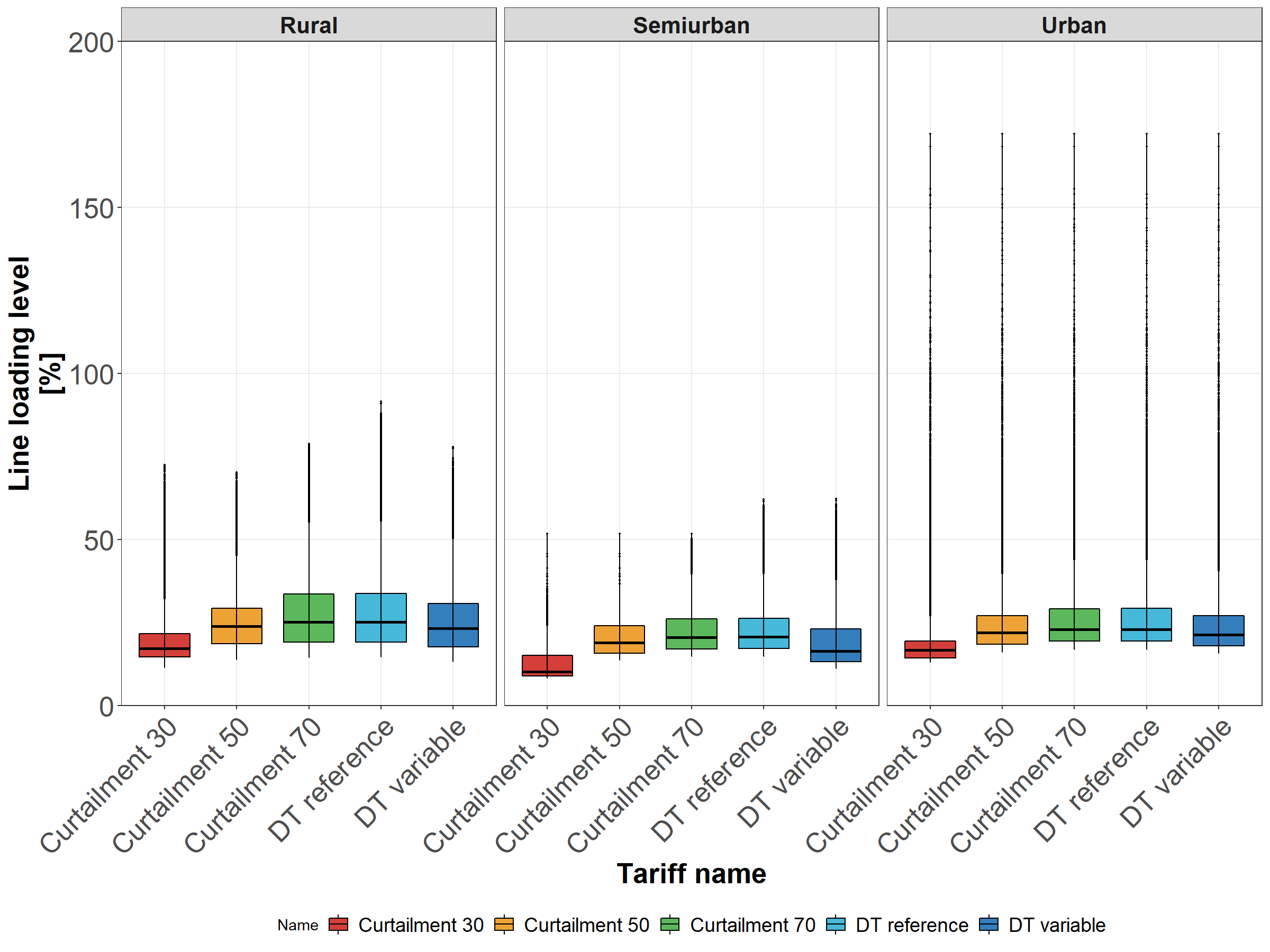}
    \caption{Line loading level distribution ($95^{th}$ percentile) under a resource sharing scenario with an optimized PV and battery design; Box plots show the median (horizontal line) and the IQR (box outline). The whiskers extend from the hinge to the highest and lowest values that are within 1.5 × IQR of the hinge, and the points represent the outliers.}
    \label{fig:lineloadP2P}
\end{figure}

\begin{figure}[H]
    \centering
    \includegraphics[scale=0.35]{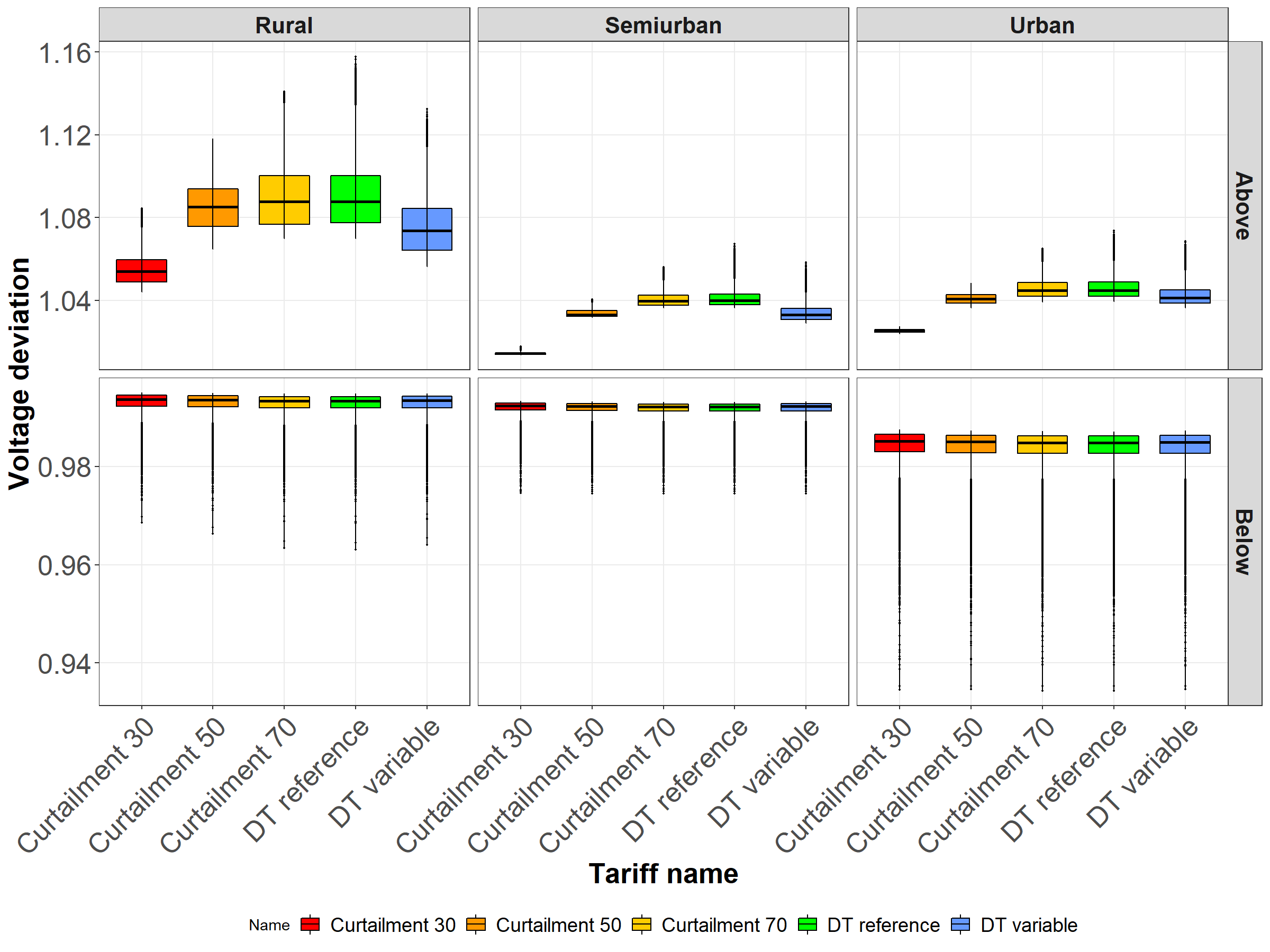}
    \caption{Voltage deviation distribution ($95^{th}$ percentile) under a resource sharing scenario with an optimized PV and battery design; Box plots show the median (horizontal line) and the IQR (box outline). The whiskers extend from the hinge to the highest and lowest values that are within 1.5 × IQR of the hinge, and the points represent the outliers.}
    \label{fig:voltP2P}
\end{figure}

\end{document}